\documentclass[twocolumn,showpacs,preprintnumbers,amsmath,amssymb,
showkeys,floatfix,aps]{revtex4}
\usepackage{graphicx}
\usepackage{dcolumn}
\usepackage{bm}
\usepackage{bigints}
\usepackage{natbib}
\usepackage{amsmath}

\begin{document}

\title{ Gaussianity test of Planck CMB polarization data using the statistics of unpolarized points (Part I)}

\author{Dmitry I. Novikov}
\affiliation{Astro-Space Center of P.N. Lebedev Physical Institute, Profsoyusnaya 84/32, Moscow, Russia 117997}
\author{Kirill O. Parfenov}
\affiliation{Astro-Space Center of P.N. Lebedev Physical Institute, Profsoyusnaya 84/32, Moscow, Russia 117997}

Corresponding author E-mail: dinovikov27@gmail.com
\begin{abstract}

  We present a detailed test for Gaussianity of Planck polarization data
  using statistics of unpolarized points on the sky, i.e. such points where
  the linear polarization vanishes. The algorithm we propose for finding such
  points is stable and guarantees their 100\% detection. Our approach allows
  us to analyze the data for Gaussianity of the signal at different angular
  scales and detect areas on the polarization maps with a significant
  contribution from unremoved non-Gaussian foregrounds. We found very strong
  deviations from Gaussianity for E and B modes in the observational
  data both over the entire sky and in some specific regions of the
  celestial sphere.

\end{abstract}


\maketitle

\section{Introduction}

The impressive results of WMAP \citep{2013ApJS..208...20B,2013JCAP...07..018C} and Planck
\citep{2014A&A...571A..15P,2018arXiv180706208P,2020A&A...641A...7P}
experiments have significantly
expanded our knowledge of fundamental physical processes in the early
Universe and brought us closer to a high-precision estimation of the main
cosmological parameters.

These experiments gave us the opportunity to study celestial maps of
the Cosmic Microwave Background (CMB) anisotropy and linear polarization,
i.e. maps of temperature $T$ and Stokes parameters $Q$ and $U$.
The information contained in these maps can be naturally divided into two
parts:\\
1. The angular power spectrum $C_\ell$;\\
2. The distribution of phases.

The power spectrum provides us with the amplitudes of perturbations at
different angular scales, while the distribution of phases
explains the nature of these perturbations, in particular, Gaussian or
non-Gaussian. It is the phases (not the
amplitudes) that determine the visual picture of the fluctuations
in the sky. Two dimensional map of a band limited signal consisting
of $\ell_{max}$ spherical harmonics contains $\ell_{max}$ times more
information than the power spectrum.

The cosmological model with inflation assumes a Gaussian nature of
fluctuations. This means that the phases should be uniformly distributed and
uncorrelated. Thus, the study of the phases distribution is a test for
Gaussianity and a verification of the inflation hypothesis.
Moreover, different models of inflation robustly predict a small
non-Gaussianity parameter $f_{NL}$
\citep{1997PhRvD..56..535L,1988PhRvD..38..465O}. Therefore, provided
that the experiment
is highly sensitive, the investigation of phases behavior may allow us
(at least in principle) to refine the model of inflation. Unfortunately,
to date the sensitivity of the experiments carried out is insufficient for a
good estimate of the small parameter $f_{NL}$, as well as for the detection of
the cosmological component of the B-mode polarization due to the
small tensor-to-scalar ratio. Thus, it makes sense to conduct a statistical
analysis of the observed signal not to refine the inflation, but to
generally establish the CMB anisotropy and polarization Gaussianity and to
detect non-Gaussian foregrounds that have not been removed from the maps.

There are many different ways to establish whether the observed signal is
Gaussian or non-Gaussian for both the scalar anisotropy field $T$ and the
tensor linear polarization field $Q,U$. Starting from the classical
theoretical papers of \citep{1970Ap......6..320D},
\citep{1986ApJ...304...15B} (BBKS) and
\citep{1987MNRAS.226..655B} (BE) on the statistics of random Gaussian fields,
various powerfull methods have been developed for studying the properties of
a signal in 2D and 3D space. Almost all of these methods are related to the
investigation of phase distribution in a random field.

Among them Minkowski functionals
\citep{1990ApJ...352....1G,1998MNRAS.297..355S,1998ApJ...507...31N,2014RAA....14..625Z,2015JCAP...02..028G,2016JCAP...07..029S,2017PhLB..771...67C,2019JCAP...01..009J,2021PhRvD.103l3523K,2024MNRAS.527..756C}, 
high-order correlations \citep{1993PhRvL..71.1124L,1994ApJ...427L..71L,2020A&A...641A...9P,2001PhRvD..63f3002K,2022JCAP...03..050G,2020PhRvD.102b3521D,2016PhRvD..94h3503S,2016PhRvD..93l3511M,2016JCAP...05..055B,2016JCAP...03..029B}, kurtosis and skewness \citep{1993ApJ...408...33L}, 
statistics of local extrema (maxima, minima and saddle points)
\citep{2022JCAP...06..006K},
clusterization of peaks \citep{1999IJMPD...8..291N,1998MNRAS.296..693B,2001IJMPD..10..501K,2021MNRAS.503..815V},
the Kullback-Leibler divergence \citep{2015JCAP...06..051B},
percolation \citep{1995ApJ...444L...1N} and
neural-network approach \citep{2015JCAP...09..064N,2014JCAP...01..018N}.

In addition, there are approaches associated with 
$\mu$ distortions \citep{2018MNRAS.478..807R,2015JCAP...09..026K,2022MNRAS.515.5847R,2023PhRvD.108j3536Z} and
CMB lensing \citep{2014MNRAS.441L..16G,2019MNRAS.489.1950B}.

Of particular interest in the study of the CMB polarization Gaussianity
is the statistics of unpolarized points in the sky, that is, those special
points at which both Stokes parameters are equal to zero. A theoretical study
of singular points for two-dimensional tensor field was carried out in
works \citep{1999IJMPD...8..189D,2000A&AT...19..213D}.
It's been shown that the polarization field in the vicinity of
such points behaves in a very characteristic way depending on their type.
Similar to the singular points of a vector field (nodes, saddles and foci),
a tensor field also has the property of having singular zero points of three
different types: comets, saddles and beaks. Obviously, they differ in
appearance from such points of a vector field. The
distribution of unpolarized points by three types and the average density of
all points per unit area of the sky were calculated for the Gaussian case
\citep{1998ApJ...507...31N,1999IJMPD...8..189D}.
It's been shown that
the obtained ratios of the numbers of points of 3 different types to the total
number of points do not depend on the power spectrum. Therefore these 3
numbers are a pure indicators of Gaussianity. At the same time the total
number of unpolarized points depends on the correlation scale.
Significant deviations from the calculated
values would indicate that the signal is non-Gaussian.

The first attempt to use unpolarized point statistics as a test for
Planck polarization data was made in \citep{2021PhRvD.104b3502K}. However, in this work
polarization was treated as a vector rather than a tensor. In addition,
the algorithm used to find unpolarized points inevitably leads to the
detection of points where the polarization is small but not zero.
At the same time, the definition of the point type described in this
article did not take into account spherical geometry.

In our paper, we propose an algorithm for searching for singular points that
assumes very precise determination of their location, guarantees that all
unpolarized points on the sky will be found and completely excludes
detection of false points that are not actually points of zero polarization.
With such an algorithm we conduct a detailed test for Gaussianity of Planck
polarization data at different angular scales $\ell_{max}$. In addition,
we show that the distribution of unpolarized points on the sky is extremely
sensitive to the presence of unremoved non-Gaussian foregrounds. This allows
us to detect areas where there is a significant contribution of such
foregrounds to the overall signal. Besides, it is known that the distribution
of white noise in Planck data is not uniform due to scanning strategy.
We show that the local density of unpolarized points is a good indicator of
the signal-to-noise ratio in different parts of the map.

Studying the distribution of points of different types is not our goal yet
(although this approach is certainly very promising) and remains beyond the
scope of this article.

The  outline of this paper is as follows:
In Section II we review the theory of unpolarized point statistics,
generalize it for E and B modes and
describe in detail the algorithm for finding such points on a pixelized
sky map with band limited $Q$ and $U$ functions. In Section III we perform a
detailed test for the Gaussianity of polarization maps using Planck
observations. To demonstrate an example of a non-Gaussian field, we first use
the shape of the Earth surface as the source of the scalar polarization
field E. We then examine the unpolarized points statistics separately for the
E and B polarization maps (SMICA) at all angular scales and identify regions
on the sky with a significant contribution from unremoved non-Gaussian
foregrounds. Finally, in Section IV we summarize our results and make
brief conclusions.

\section{Singular (unpolarized) points of CMB polarization field}

  \subsection{Stokes parameters}

\begin{figure}[!htbp]
  \includegraphics[width=1\columnwidth]{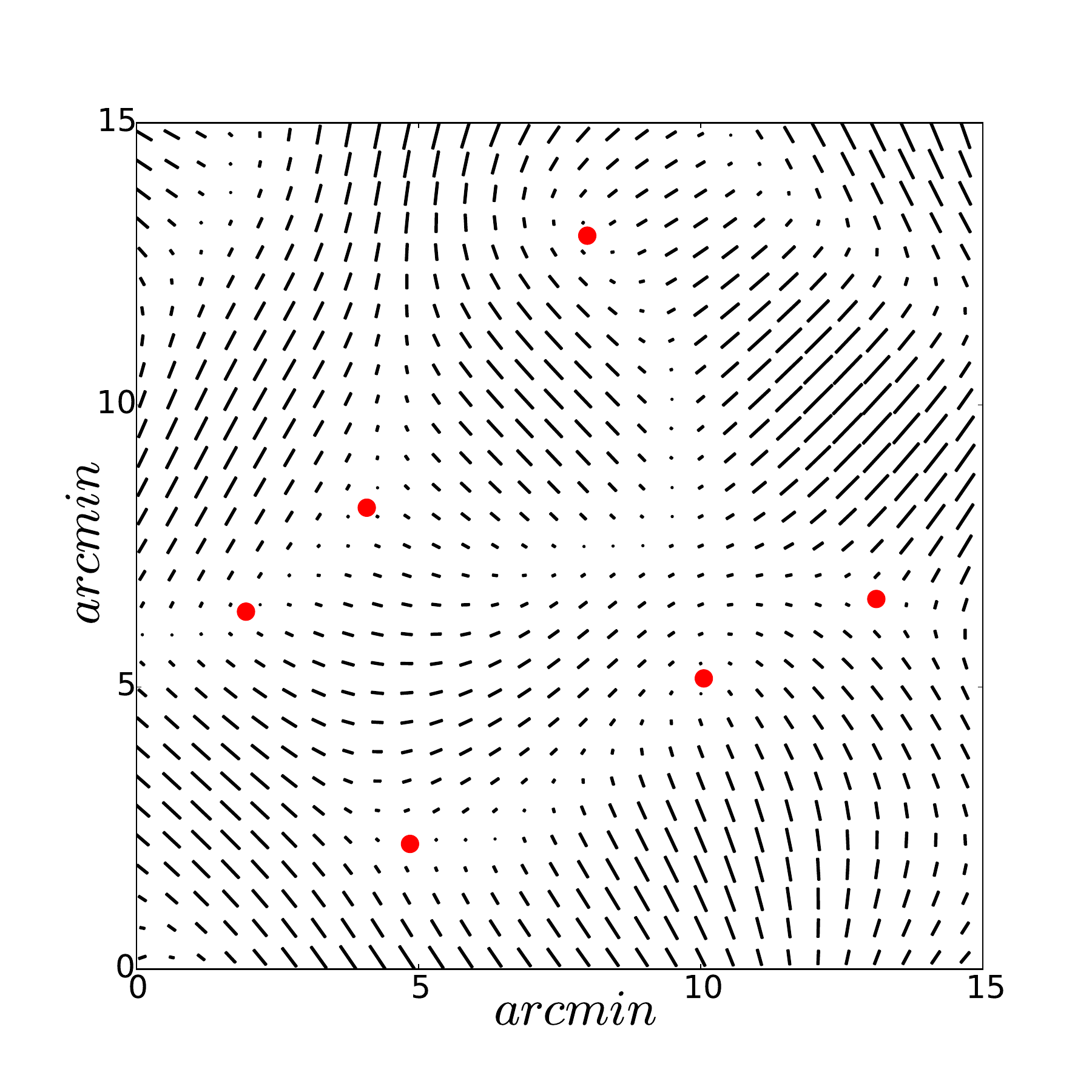}
  \caption{Polarization map for a small ($15'\times 15'$) patch of the sky with
    Planck resolution. The lengths of the segments are proportional to the
    degree of polarization and their orientations correspond to the directions
    of polarization. Red dots are singular points with zero polarization.}
\end{figure}
  
The linear polarization in the plane of the local coordinate system ($x,y$)
perpendicular to the direction of radiation propagation is represented by
two Stokes parameters $Q$ and $U$ and can be completely described in terms of
the scalar $E(x,y)$ and the pseudoscalar $B(x,y)$ fields:
\begin{equation}
  \begin{array}{l}
    \vspace{0.5cm}
    Q=\left[\frac{\partial^2}{\partial x^2}-\frac{\partial^2}{\partial y^2}
    \right]E
    -2\frac{\partial ^2}{\partial x\partial y}B,\\
    
      U=2\frac{\partial ^2}{\partial x\partial y}E+\left[\frac{\partial^2}
      {\partial x^2}-\frac{\partial^2}{\partial y^2}\right]B
    \end{array}
  \end{equation}
Two-dimensional functions $E$ and $B$ correspond to the 'electric' and
'magnetic' modes of polarization. In the spherical coordinate system
$\theta,\varphi$ the differential operators in Eq. (1) look as follows:

\begin{equation}
  \begin{array}{l}
    \vspace{0.5cm}
    \frac{\partial^2}{\partial x^2}-\frac{\partial^2}{\partial y^2}=
    \frac{\partial^2}{\partial\theta^2}-\frac{\cos\theta}{\sin\theta}
    \frac{\partial}{\partial\theta}
    -\frac{1}{\sin^2\theta}\frac{\partial^2}{\partial\varphi^2} ,\\
    
    \frac{\partial ^2}{\partial x\partial y}=\frac{1}{\sin^2\theta}
    \frac{\partial}{\partial\varphi}\left[\sin\theta\frac{\partial}
      {\partial\theta}-\cos\theta\right],
    \end{array}
\end{equation}
where the x coordinate is co-directed with $\theta$ along the meridian and
y is perpendicular to x and directed along the great circle. 
Thus, band limited continuous functions $Q$ and $U$ can be obtained using
band limited functions $E$ and $B$ defined on the sphere by means of
spherical harmonics:
\begin{equation}
  \begin{array}{l}
    E,B=\sum\limits_{\ell=2}^{\ell_{{}_{max}}}\sum\limits_{m=-\ell}^{\ell}
    a_{\ell m}^{E,B}\cdot Y_{\ell m}(\theta,\varphi).
    \end{array}
\end{equation}
If the Stokes parameters are generated by random Gaussian fields $E$
and $B$, then all statistical characteristics of the polarization 
are determined by the power spectrum $C_\ell=C_\ell^E+C_\ell^B$:
\begin{equation}
    \begin{array}{l}
   C_\ell=\frac{1}{2\ell+1}\sum\limits_{m=-\ell}^\ell\left[
     \left(a_{\ell m}^{E}\right)\left(a_{\ell m}^{E}\right)^*+
     \left(a_{\ell m}^{B}\right)\left(a_{\ell m}^{B}\right)^*\right],
   \end{array}
 \end{equation}
where the asterisk denotes the complex conjugate.
Equivalent to Eqs. (1-3), the Stokes parameters on the celestial sphere
can also be written using spin-weighted spherical harmonics
${}_{\pm 2}Y_{\ell m}$ \citep{PhysRevD.55.1830}.
Linear polarization is a two-dimensional tensor and can be described at each
point on the sphere by the degree of polarization $P$ and its orientation
$\phi$ relative to the chosen coordinate system:
  \begin{equation}
    \begin{array}{l}
     P^2=Q^2+U^2,\hspace{1cm}\tan(2\phi)=U/Q. 
     \end{array} 
  \end{equation}

  Visually, the polarization map can be represented as in Fig. 1, where the
  lengths of the segments are proportional to the degree of polarization, and
  their orientations correspond to the direction of linear polarization at a
  given point on the map. In the general case the polarization statistics
  is determined by the power spectrum $C_\ell$ and the phase distribution,
  that is, possible specific correlations between the coefficients
  $a_{\ell m}^{E,B}$.
  
  The mean square of the degree of polarization can be written as follows:
\begin{equation}
  \begin{array}{l}
    \vspace{0.3cm}
    \langle P^2\rangle=\langle Q^2\rangle+\langle U^2\rangle=2\sigma_0^2,\\
    \sigma_0^2=\sum\limits_{l=2}^{\ell_{max}}(2\ell+1)
    \frac{(\ell+2)!}{(\ell-2)!}(C_\ell^E+C_\ell^B),
   \end{array}
\end{equation}
where $\langle\rangle$ means the ensemble average over the entire sky.
For further research, we will also need derivatives of the Stokes parameters with respect to the variables $x,y$. The mean squares for these quantities are as follows:

\begin{equation}
  \begin{array}{l}
    \vspace{0.3cm}
    \langle Q_x^2\rangle+\langle Q_y^2\rangle=
    \langle U_x^2\rangle+\langle U_y^2\rangle=\sigma_1^2,\\
    \sigma_1^2=\sum\limits_{l=2}^{\ell_{max}}(2\ell+1)
    \frac{(\ell+3)!}{(\ell-3)!}
    \left(C_\ell^E+C_\ell^B\right).
   \end{array}
\end{equation}

Below, for convenience, we will use the following notations for normalized dimensionless values:
\begin{equation}
   \begin{array}{l}
     \vspace{0.2cm}
     q=\frac{Q}{\sigma_0},\hspace{0.2cm}u=\frac{U}{\sigma_0},\hspace{0.2cm}
     p=\frac{P}{\sigma_0}\\
     q_x=\frac{Q_x}{\sigma_1},\hspace{0.2cm}q_y=\frac{Q_y}{\sigma_1},
     \hspace{0.2cm}
     u_x=\frac{U_x}{\sigma_1},\hspace{0.2cm}u_y=\frac{U_y}{\sigma_1}.
   \end{array}
 \end{equation}
 
  \subsection{Number density of unpolarized points}

Let us consider a small region of the celestial sphere where the geometry
is approximately flat. Having fixed the coordinate system
$x,y$ one can find isocontour lines $q(x,y)=0$ and $u(x,y)=0$ for the random
fields $q$ and $u$. These
lines intersect each other at some points $x_0,y_0$ where both Stokes
parameters are zero and, therefore, there is no linear
polarization at such points $p(x_0,y_0)=0$ (see Fig. 1). As can be seen in this
Figure, the polarization structure in the vicinity of such points is different
and is determined by the type of point (saddle, comet or beak
\citep{1999IJMPD...8..189D}). In this paper,
we are interested in the general statistics of unpolarized points without
dividing them into different types.
The behavior of the Stokes parameters in the vicinity of $x_0,y_0$ is determined
by their first derivatives at this point:
\begin{figure*}[!htbp]
  \includegraphics[width=0.49\textwidth]{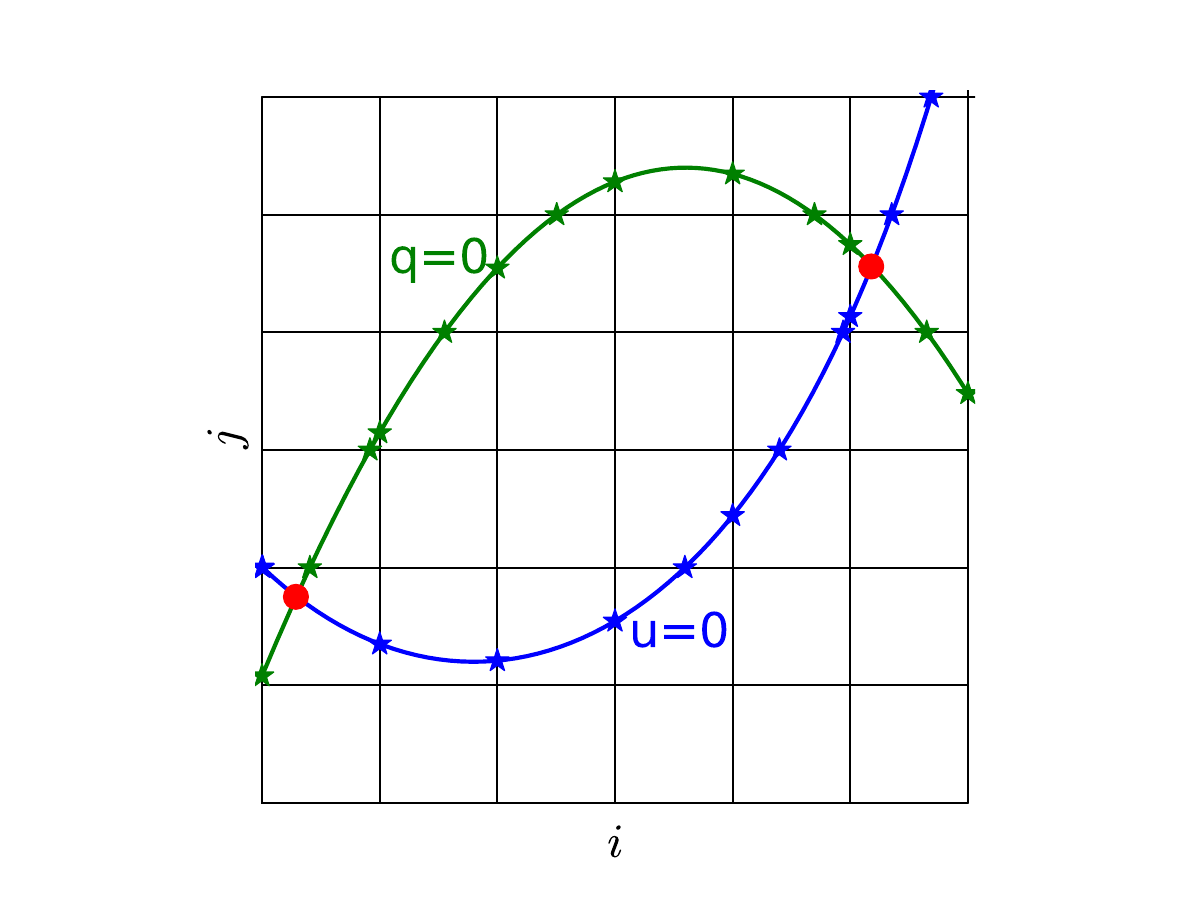}
  \includegraphics[width=0.49\textwidth]{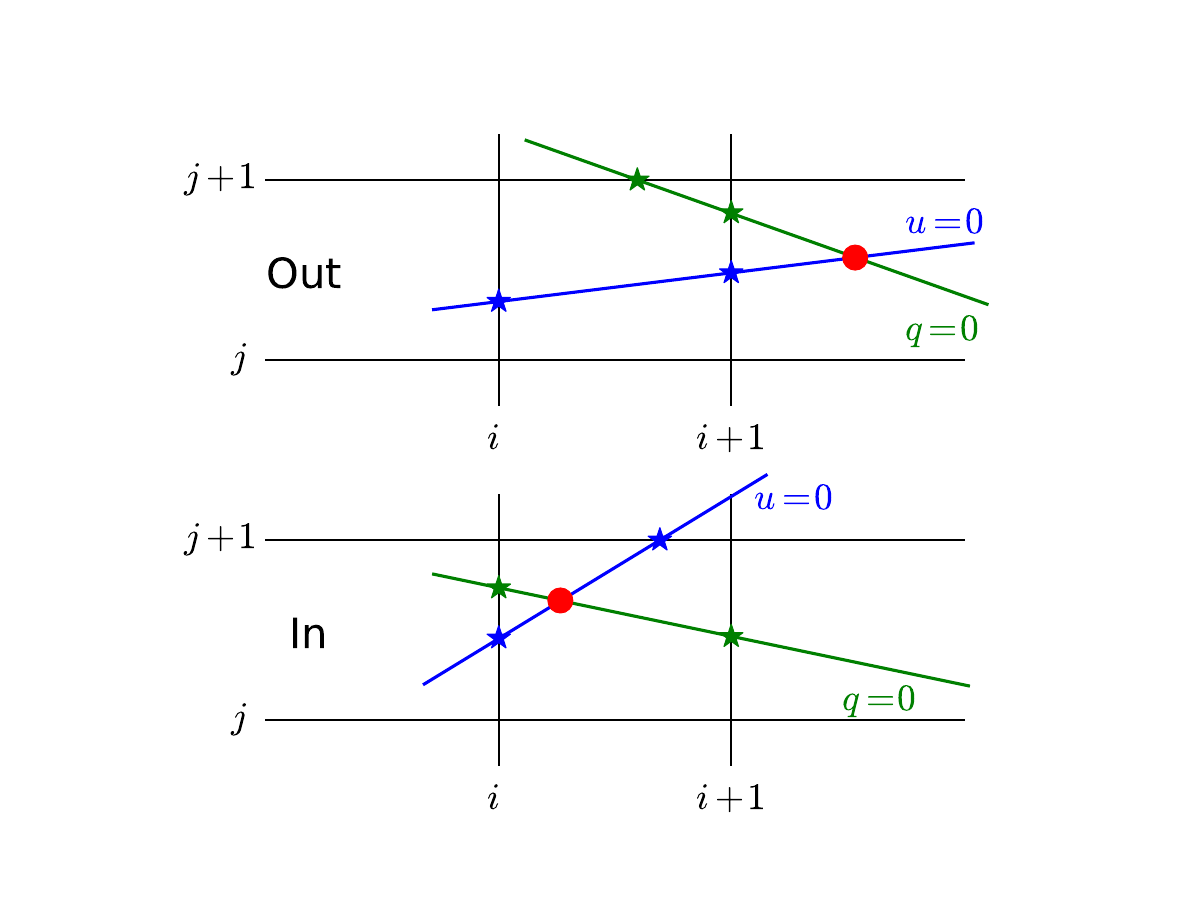}
  \caption{The algorithm for finding unpolarized points. {\it Left panel:}
    Green and blue stars correspond to the intersections of the
    lines $q=0$ and $u=0$ with the grid lines. Unpolarized points are marked
    in red. {\it Right panel:} A way to distinguish the
    intersection point of the lines $q=0$ and $u=0$ inside the square
    $(i,j),(i+1,j),(i,j+1),(i+1,j+1)$ from a point outside this square.
    Only a point that falls within this square is considered unpolarized.
    This ensures that the same point will not be counted more than once.}
\end{figure*}
 \begin{equation}
  \begin{array}{l}
    \begin{pmatrix}q\\u\end{pmatrix}=\frac{\sigma_1}{\sigma_0}\cdot
    \begin{pmatrix}q_x&q_y\\u_x&u_y\end{pmatrix}\cdot
     \begin{pmatrix}\Delta x\\ \Delta y\end{pmatrix},
   \end{array}
\end{equation}
 where $\Delta x=x-x_0$, $\Delta y=y-y_0$. Knowing the joint
 probability distribution function $\Phi$ of random values
 $q,u,q_x,q_y,u_x,u_y$ one can predict the density of unpolarized points
 per unit area of the celestial sphere. For the Gaussian $E$ and $B$ we have:
 \begin{equation}
  \begin{array}{l}
    \langle q^2\rangle=\langle u^2\rangle=1\\
    \langle q_x^2\rangle=\langle q_y^2\rangle=
    \langle u_x^2\rangle=\langle u_y^2\rangle=\frac{1}{2},
   \end{array}
 \end{equation}
 and there is no correlations between these six random values.
 The relations in Eq. (10) remain valid both separately for scalar and pseudoscalar
 fields, and for any combination of them if E and
 B are uncorrelated (which is the case for the standard cosmological model).
 Thus, the function $\Phi(q,u,q_x,q_y,u_x,u_y)$ looks as follows:
 \begin{equation}
   \begin{array}{l}
     \vspace{0.3cm}
     d\Phi=\frac{1}{2\pi^3}\cdot
     e^{-G}\cdot dqdudq_xdq_ydu_xdu_y,\\
     G=\frac{q^2}{2}+\frac{u^2}{2}+
     q_x^2+q_y^2+u_x^2+u_y^2.
   \end{array}
 \end{equation}
 Similar to \citep{1986ApJ...304...15B} (BBKS) and \citep{1987MNRAS.226..655B} (BE)
 (where the statistics of local extrema in
 three-dimensional and two-dimensional random scalar fields was found),
 we substitute $q=u=0$ and
 $dqdu=\sigma^2_1/\sigma^2_0|q_xu_y-u_xq_y|dxdy$ into
 Eq. (11). Therefore, the average number density of unpolarized points
 $\langle n\rangle$ can be written in the following form:
  \begin{equation}
   \begin{array}{l}
     \vspace{0.2cm}
  \langle n\rangle=\frac{1}{2\pi^3}\frac{\sigma_1^2}{\sigma_0^2}
     \int e^{-g}\mid d\mid dq_xdq_ydu_xdu_y,\\
      \vspace{0.2cm}
      g=q_x^2+q_y^2+u_x^2+u_y^2,\\
      d=q_xu_y-u_xq_y.
   \end{array}
 \end{equation}
 Integrating over the entire possible range for the values $q_x,q_y,u_x,u_y$
  from $-\infty$ to $+\infty$ gives us the total expected number of
 such points on the sky:
 \begin{equation}
     \langle N_p\rangle=A\cdot\langle n\rangle=\frac{A}{4\pi r_c^2},
 \end{equation}
 where $r_c=\frac{\sigma_0}{\sigma_1}$ is the correlation scale and $A$
 is the area of the sky region under concideration
 ($A=4\pi$ for the whole sky).
 
 Note that Eq. (13) is true only in the flat approximation
 ($\ell_{max}\gg 1$) and provided that the sum 
\begin{equation}
  \sigma_1^2\sim\sum\limits_{\ell=2}^{\infty}\ell^{7}C_\ell
  \rightarrow\infty
 \end{equation}
diverges, i.e. if, for example, the spectrum has a power law form
$C_\ell\sim\ell^{{}^\gamma}$, then $\gamma$ should satisfy $\gamma\ge -8$.
The condition in Eq. (14) ensures a decrease in the
correlation radius and, consequently, an increase in the number of
unpolarized points with increasing $\ell_{max}$, see Eq. (6,7,13).

For small $\ell_{max}$,
 that is, if the correlation scale is comparable to the curvature
 radius, then $\langle N_p\rangle$ must be found
 numerically for a given power spectrum.
 
Unpolarized points in the Gaussian case are not uniformly distributed. There
is a two-point autocorrelation function describing the statistics of their
distribution, which is determined by the spectrum $C_\ell$
(similar to the clusterization of peaks in random Gaussian fields
\citep{1986ApJ...304...15B} ).
However, at angular scales greater than the correlation radius, their
distribution can be considered as homogeneous and isotropic.

Counting the total number of unpolarized points makes it possible to test
the polarization of the relic radiation for Gaussianity at different angular
scales. The polarization map is determined by harmonics from $\ell=2$ to
$\ell=\ell_{max}$, Eq. (3). Varying $\ell_{max}$ from $\ell_{max}=2$ to Planck
resolution ($\ell=2048$) one can study the dependence of the number
of zero polarization points on $\ell_{max}$: $N_p=N_p(\ell_{max})$.
In the case of Gaussianity this function should not deviate from the expected mean much more than the possible variations estimated as
$N_p-\langle N_p\rangle\sim\sqrt{\langle N_p\rangle}$. In our studies
we numerically find the average possible positive and negative deviations
from $\langle N_p\rangle$.

It is easy to see, that $N_p\sim\ell_{max}^2$,
(unless the spectrum has some weird discontinuity). Any features of the spectrum (for example, acoustic oscillations)
inevitably manifest themselves in the behaviour of $N_p(\ell_{max})$.

The approach described above is a general test for Gaussianity when examining
the entire sky (except for areas that have been masked out).
Even more interesting
is the study of the spatial distribution of unpolarized points on the sky.
This makes it possible to detect local non-Gaussianities, i.e. areas where
the observed signal contains unremoved foregrounds, systematic errors or
variations in white noise due to Planck's extremely anisotropic sky scanning.

\subsection{Unpolarized points on a pixelized map}

Finding unpolarized points on a pixelized sky is not an easy task due to the
discrete nature of the Stokes parameter maps available for study.
The method for searching such points should:\\
$\bullet$ exclude the possibility of detecting false points that are not actually
unpolarized;\\
$\bullet$ guarantee that all unpolarized points will be detected;\\
$\bullet$ ensure high accuracy of finding the location of each point on the map.

The algorithm for searching for unpolarized points is demonstrated in Fig. 2.

For simplicity and convenience we use the
equidistant pixelization along parallels and
meridians: $\theta_i=hi$, $\varphi_j=hj$, where $h$ is the
angular step.
Thus the values of the continuous band limited functions $q$
and $u$ are known only at the intersections of these grid lines:
$q(\theta_i,\varphi_j)=q_i^j$, $u(\theta_i,\varphi_j)=u_i^j$.
In order to find points that belong to, for example, an isocontour line
$q=0$, each pair of adjacent pixels is checked for the condition of changing the sign of $q$ along the grid lines. If such a condition is met, then the
line $q=0$ intersects the grid
line between two pixels located next to each other. The location of the
intersection point can be found using the linear interpolation.

Thus, the zero points $\theta_q,\varphi_q$ for $q$ parameter along,
for example, the grid line $\varphi_j$ are found as follows:
 \begin{equation}
   \begin{array}{l}
     \vspace{0.2cm}
     if\hspace{0.5cm} q_i^j\cdot q_{i+1}^j\le 0,\hspace{0.5cm}then\\
     \theta_q=\theta_i+h\frac{q_i^j}{q_i^j-q_{i+1}^j},\hspace{0.5cm}
     \varphi_q=\varphi_j.     
   \end{array}
 \end{equation}

The same should be done for the grid lines $\theta_i$.
The zero isocontour line points $\theta_u,\varphi_u$ for the parameter
$u$ are found in a similar way (Fig. 2, left panel).

The remaining part of the problem is to find the intersection of the lines
$q=0$ and $u=0$. To do this, each quadruple of points
$(i,j), (i+1,j), (i,j+1), (i+1,j+1)$ is checked. If on the segments connecting
these points there are two zero points for $q$ parameter
($\theta_q^{1,2},\varphi_q^{1,2}$) and two zero points for $u$ parameter
($\theta_u^{1,2}\varphi_u^{1,2}$),
then two straight lines passing through these pairs of points intersect each other at the place where $q=u=0$. It is obvious that to find the intersections
of lines we must use spherical geometry. In order to avoid detecting the same
unpolarized point several times we consider the intersection point to be the
point of zero polarization only if it is inside the square bounded by 4
points indicated above (Fig. 2, right panel).

In fact, we don't need to find all the lines $q=0$ and $u=0$. It is enough to
check each quadruple of neighboring pixels for the presence of two pairs of
points ($\theta_q^{1,2},\varphi_q^{1,2}$) and ($\theta_u^{1,2},\varphi_u^{1,2}$).
After that we should check if the intersection of the lines connecting these
pairs of points is inside the square $(i,j), (i+1,j), (i,j+1), (i+1,j+1)$.

The algorithm described above is quite fast, since to implement it, we need
to visit each pixel only once. However, for this method to work reliably, it
is necessary that the map resolution is high enough, i.e. the relation
$h\ll r_c$ is to be satisfied. Empirically, we came to the
conclusion that the number of pixels along the $\varphi$ coordinate
(which in our case is twice the number of pixels along $\theta$) 
should exceed $\ell_{max}$ by at least 16 times.
\section{Gaussianity test for Planck polarization data.}

One of the ideas of our approach is to compare the observed signal with its
spectrum and phases with a signal that has an identical spectrum but random,
uncorrelated and uniformly distributed phases. Thus, we compare two signals:
the observed one and a Gaussian one with the same power spectra.
\begin{figure*}[!htbp]
  \includegraphics[width=0.49\textwidth]{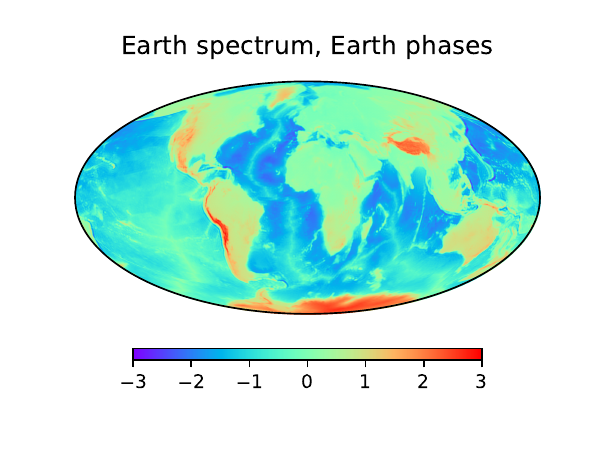}
  \includegraphics[width=0.49\textwidth]{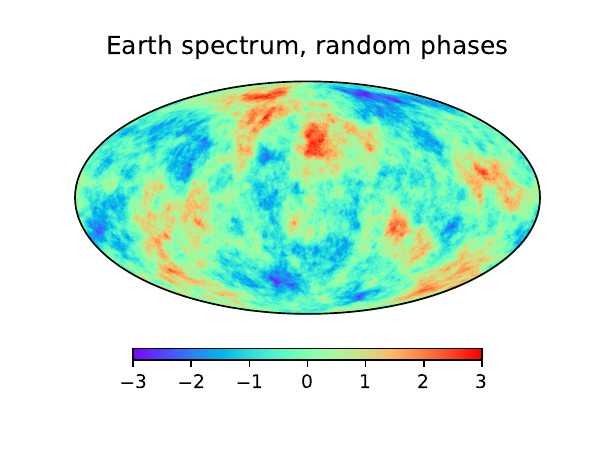}
  \includegraphics[width=0.49\textwidth]{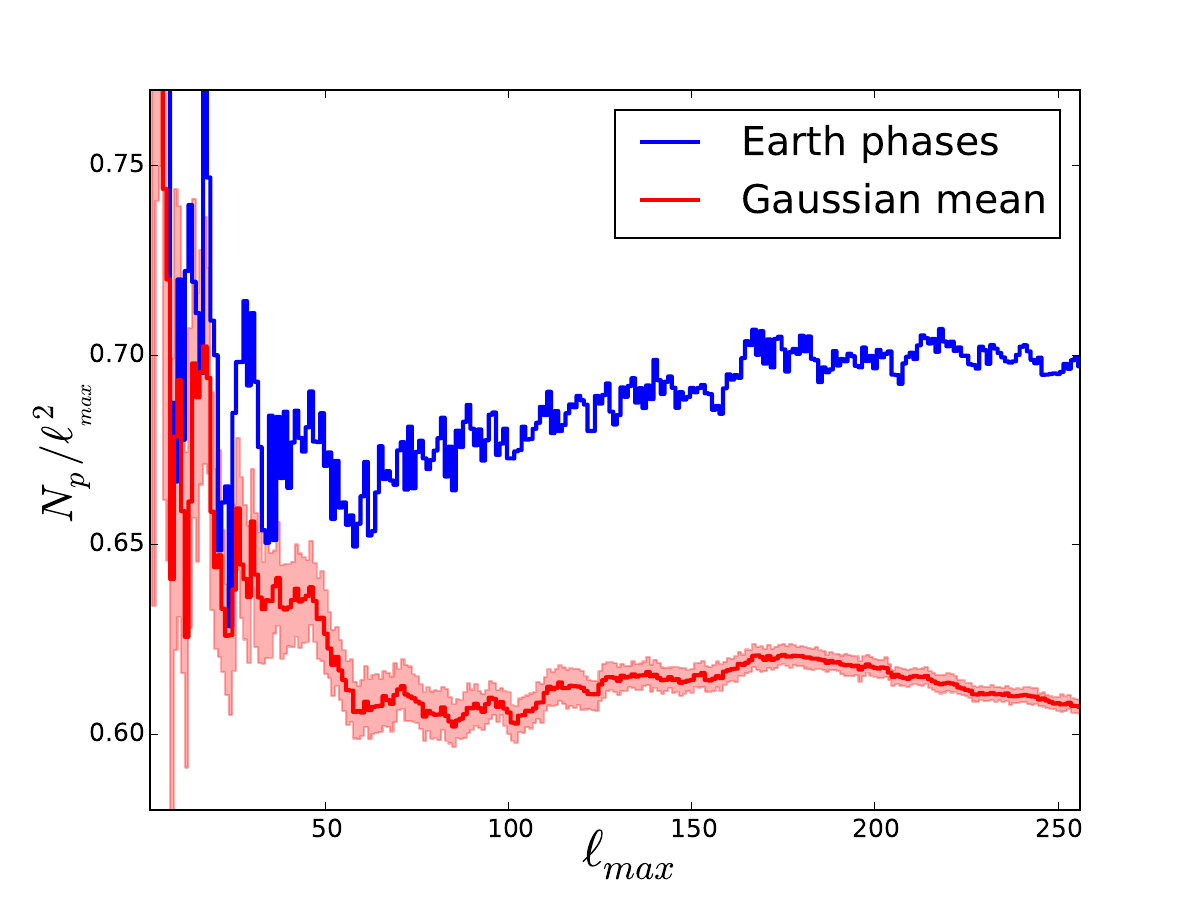}
  \includegraphics[width=0.49\textwidth]{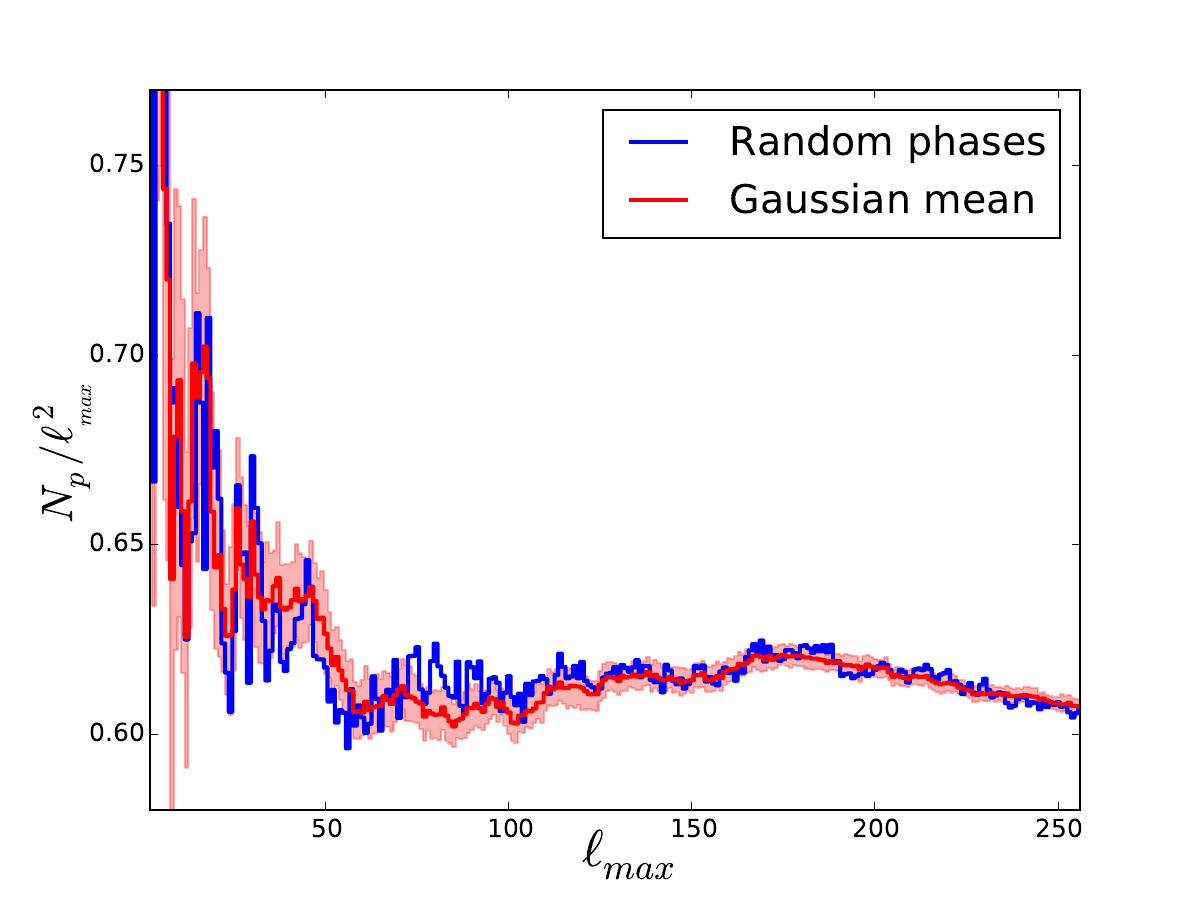}
  \caption{ {\it Left panel (top)}:
    The scalar field (map) of the Earth
    surface as a source of E-polarization.
    The color scale is given in units of
    $\sigma$ deviation from the mean. {\it Left panel (bottom)}:
    The function $N_p(\ell_{max})$ as a dependence of the number of
    unpolarized points on the number of harmonics $\ell_{max}$.
    The red line corresponds to the ensemble average
    $\langle N_p(\ell_{max})\rangle$ of the number of zero points for a
    Gaussian polarization field with the
    Earth spectrum $C_\ell^{{}^{\boldsymbol{\oplus}}}$.
    The shaded area corresponds to 1 $\sigma$
    deviations from $\langle N_p(\ell_{max})\rangle$ in the positive and
    negative directions. The blue line is the number of unpolarized
    points for the field with the Earth spectrum and Earth phases.
    {\it Right panel (top)}: The scalar field with Earth spectrum but random
    phases. {\it Right panel (bottom)}: The same as the
    {\it left panel (bottom)}, but the blue line shows the number of
    unpolarized points for the scalar Gaussian field shown in the
    {\it right top} panel.
    }
\end{figure*}
In this Section we first demonstrate an example of $E$ polarization due to a
non-Gaussian scalar field in the form of the surface shape of the planet
Earth. After that we proceed to test the Planck data separately for the
E and B components using SMICA CMB maps from the Planck DR3 release.
\subsection{The surface of the planet Earth}
Since the Earth, like the sky, is a sphere, the surface of our planet can be
expressed in spherical harmonics as a scalar field
$E^{{}^{\boldsymbol{\oplus}}}$. The Stokes parameters $Q^{{}^{\boldsymbol{\oplus}}}$ and
$U^{{}^{\boldsymbol{\oplus}}}$ generated by such a field are found using Eq. (1):
\begin{equation}
   \begin{array}{l}
     \vspace{0.2cm}
     E^{{}^{\boldsymbol{\oplus}}}=\sum\limits_{\ell m}a_{\ell m}^{{}^{\boldsymbol{\oplus}}}
     \cdot Y_{\ell m}(\boldsymbol{\eta}),
     \hspace{0.4cm}\boldsymbol{\eta}=(\theta,\varphi),\\
     Q^{{}^{\boldsymbol{\oplus}}}=\left[\frac{\partial^2}{\partial x^2}-
       \frac{\partial^2}{\partial y^2}
       \right]E^{{}^{\boldsymbol{\oplus}}},\hspace{0.5cm}
     U^{{}^{\boldsymbol{\oplus}}}=2\frac{\partial ^2}{\partial x\partial y}
     E^{{}^{\boldsymbol{\oplus}}}.
   \end{array}
\end{equation}

In terms of
spin-weighted harmonics, the polarization parameters are as follows:
\begin{equation}
  \begin{array}{l}
    \vspace{0.3cm}
    Q^{{}^{\boldsymbol{\oplus}}}
    =\frac{1}{2}\sum\limits_{\ell,m}
   \sqrt{\frac{(\ell+2)!}{(\ell-2)!}}\cdot a_{\ell m}^{{}^{\boldsymbol{\oplus}}}
    \left[{}_{2}Y_{\ell m}(\boldsymbol{\eta})+
      {}_{-2}Y_{\ell m}(\boldsymbol{\eta})\right], \\
    U^{{}^{\boldsymbol{\oplus}}}
    =\frac{1}{2i}\sum\limits_{\ell,m}
    \sqrt{\frac{(\ell+2)!}{(\ell-2)!}}\cdot a_{\ell m}^{{}^{\boldsymbol{\oplus}}}
    \left[{}_{2}Y_{\ell m}(\boldsymbol{\eta})-
         {}_{-2}Y_{\ell m}(\boldsymbol{\eta})\right].
    \end{array}
\end{equation}
At the same time, for a random Gaussian field
$\widetilde{E}^{{}^{\boldsymbol{\oplus}}}$ with a spectrum identical to the spectrum
of the field $E^{{}^{\boldsymbol{\oplus}}}$, the Stokes parameters
correspond to the coefficients
$\widetilde{a}_{\ell m}^{{}^{\boldsymbol{\oplus}}}$:
\begin{equation}
  \begin{array}{l}
    \vspace{0.3cm}
    \widetilde{Q}^{{}^{\boldsymbol{\oplus}}}
    =\frac{1}{2}\sum\limits_{\ell,m}
    \sqrt{\frac{(\ell+2)!}{(\ell-2)!}}\cdot\widetilde{a}_{\ell m}^{{}^{\boldsymbol{\oplus}}}
    \left[{}_{2}Y_{\ell m}(\boldsymbol{\eta})+
      {}_{-2}Y_{\ell m}(\boldsymbol{\eta})\right], \\
    \widetilde{U}^{{}^{\boldsymbol{\oplus}}}
    =\frac{1}{2i}\sum\limits_{\ell,m}
    \sqrt{\frac{(\ell+2)!}{(\ell-2)!}}\cdot\widetilde{a}_{\ell m}^{{}^{\boldsymbol{\oplus}}}
    \left[{}_{2}Y_{\ell m}(\boldsymbol{\eta})-
         {}_{-2}Y_{\ell m}(\boldsymbol{\eta})\right],
    \end{array}
\end{equation}
where $\widetilde{a}_{\ell m}^{{}^{\boldsymbol{\oplus}}}$ are Gaussian random
numbers with zero mean, equal variances and normalized in such a way, that:
\begin{equation}
  \begin{array}{l}
    \sum\limits_{m=-\ell}^{\ell}
    \left(\widetilde{a}_{\ell m}^{{}^{\boldsymbol{\oplus}}}\right)\cdot
    \left(\widetilde{a}_{\ell m}^{{}^{\boldsymbol{\oplus}}}\right)^*=
    \sum\limits_{m=-\ell}^{\ell}
    \left(a_{\ell m}^{{}^{\boldsymbol{\oplus}}}\right)\cdot
    \left(a_{\ell m}^{{}^{\boldsymbol{\oplus}}}\right)^*.
    \end{array}
\end{equation}
The equality in Eq. (19) ensures that the polarization spectra
in Eqs. (17,18) are identical:
$\widetilde{C}_\ell^{{}^{\boldsymbol{\oplus}}}=C_\ell^{{}^{\boldsymbol{\oplus}}}$.

The maximum number of harmonics $\ell_{max}$ can be varied to study the
Gaussianity of
the signal at different angular scales. Using the algorithm described in the
previous Section, we find the number of unpolarized points on the maps by
successively increasing $\ell_{max}$ from $\ell_{max}=2$ to the maximum
resolution. In our case for the Earth we used the maximum resolution
$\ell_{max}=256$, which corresponds to approximately 50 miles.
Thus, we construct the function $N_p(\ell_{max})$ as a dependence of the
total number of unpolarized points on the number of harmonics used.
The more harmonics make up the map, the smaller the correlation radius
becomes and the more unpolarized points appear.

In Fig. 3 we show the results obtained for a
polarization map with the Earth spectrum and Earth phases and
for a Gaussian map, which has the same Earth spectrum but
random phases. The huge difference between them is explained by the special
phasing of the harmonics describing the Earth surface. The physics of the
planet formation is not determined by a stochastic process, has nothing to
do with the Central Limit Theorem, and does not generate Gaussian statistics
for the surface shape. This example demonstrates the sensitivity of the
unpolarized point statistics to the non-Gaussianity of the signal.

\subsection{Planck polarization data}
\begin{figure*}[!htbp]
  \includegraphics[width=0.49\textwidth]{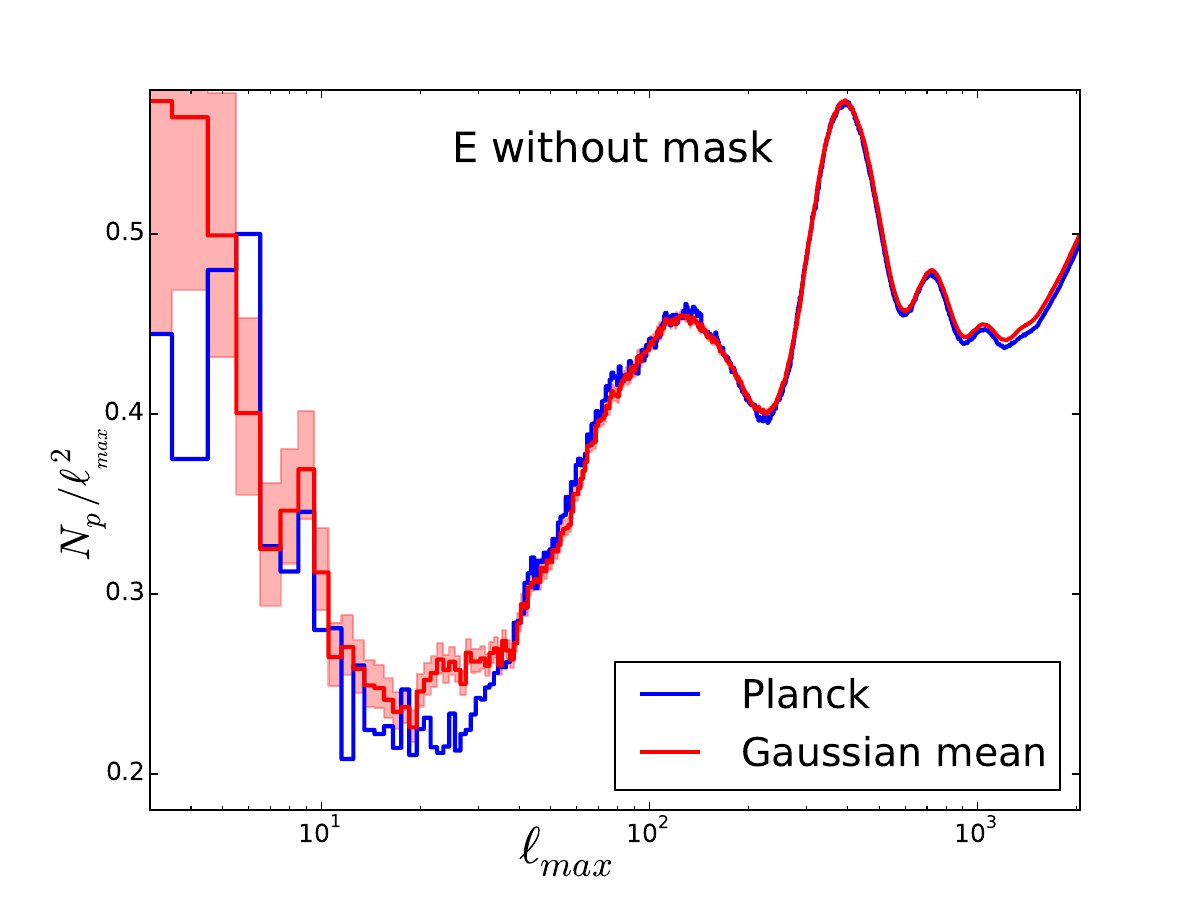}
  \includegraphics[width=0.49\textwidth]{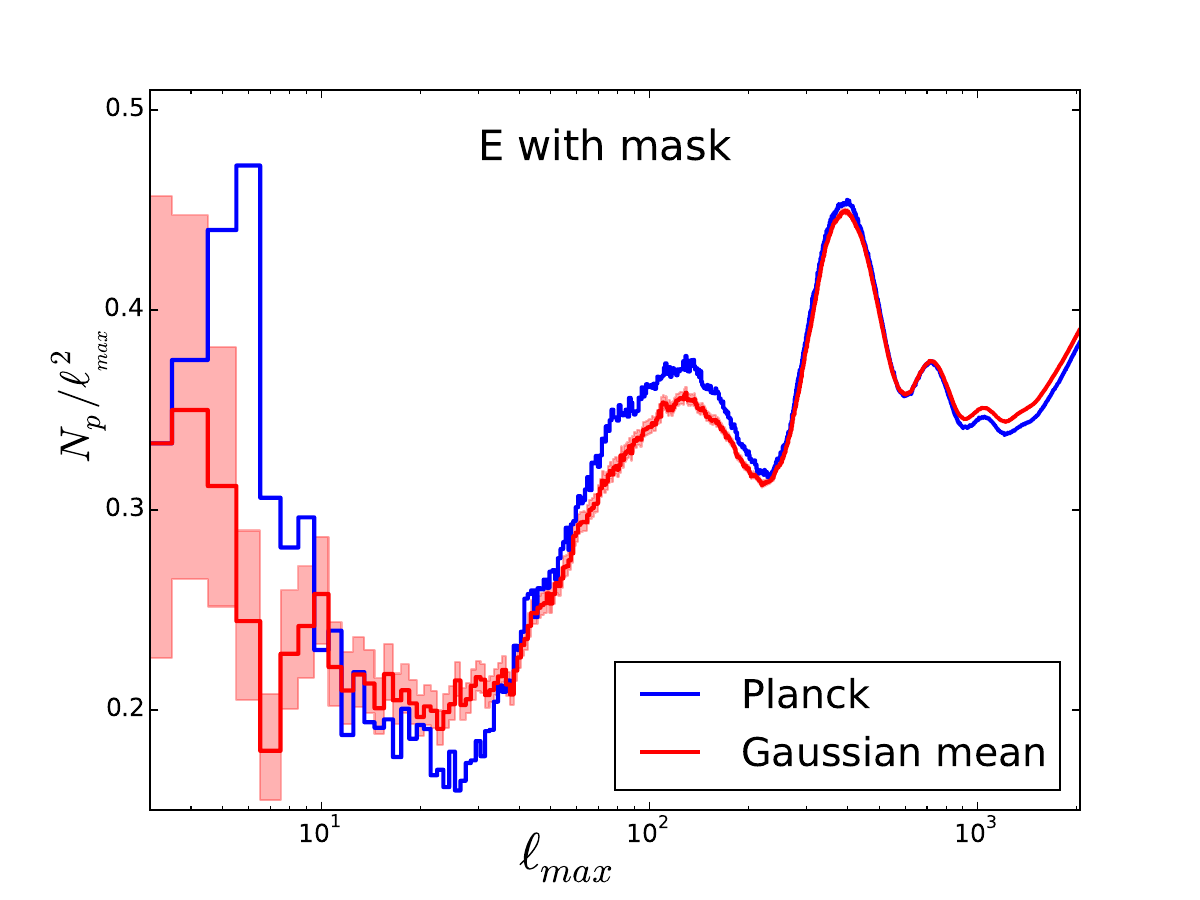}
  \includegraphics[width=0.49\textwidth]{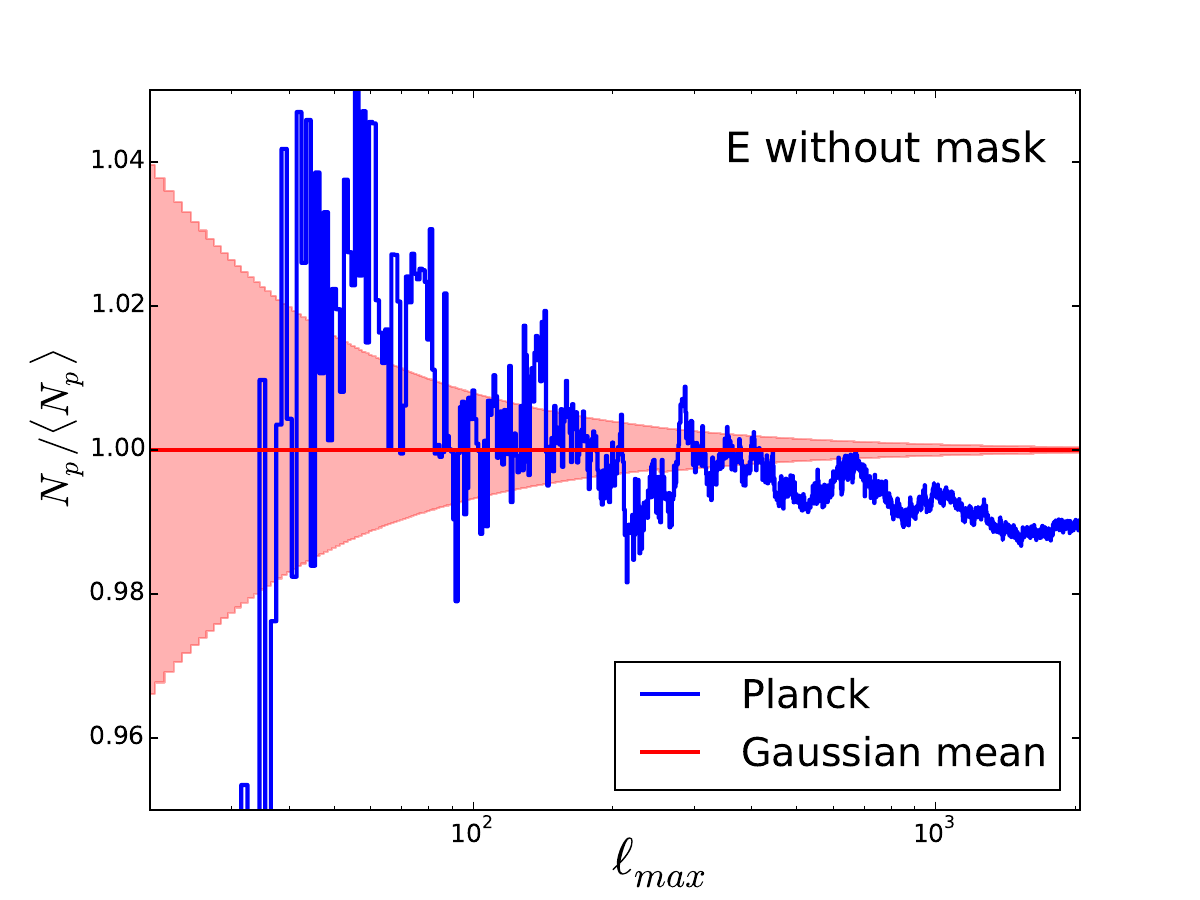}
  \includegraphics[width=0.49\textwidth]{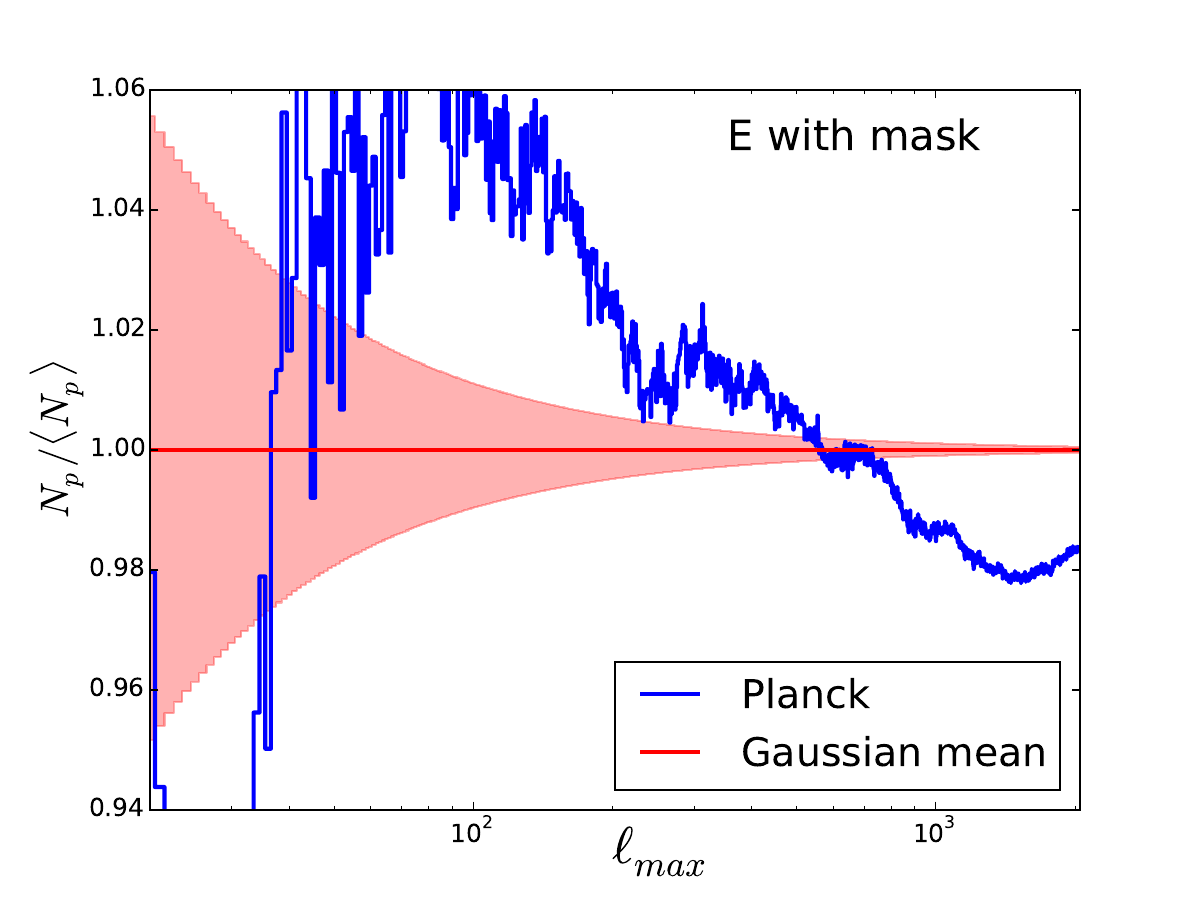}
  \caption{Gaussianity test for E polarization. {\it Upper left panel}:
    The blue line represents the number of unpolarized points on the entire
    sky as a function of the total number of harmonics used $\ell_{max}$.
    The red line is the mean expected number of such points
    $\langle N_p(\ell_{max})\rangle$ for a Gaussian
    field with the observed Planck spectrum. The shaded area corresponds to
    1 $\sigma^+(\ell_{max})$ upward and 1 $\sigma^-(\ell_{max})$ downward
    deviations from the mean. Oscillations of the function $N_p$ are a
    manifestation of acoustic oscillations during the recombination epoch.
    {\it Lower left panel}: Same as upper left, but all data are normalized
    to $\langle N_p\rangle$.
    {\it Upper right and lower right panels}: Same as left panels, but with
    the polarization confidence mask applied.}
\end{figure*}
\begin{figure*}[!htbp]
  \includegraphics[width=0.49\textwidth]{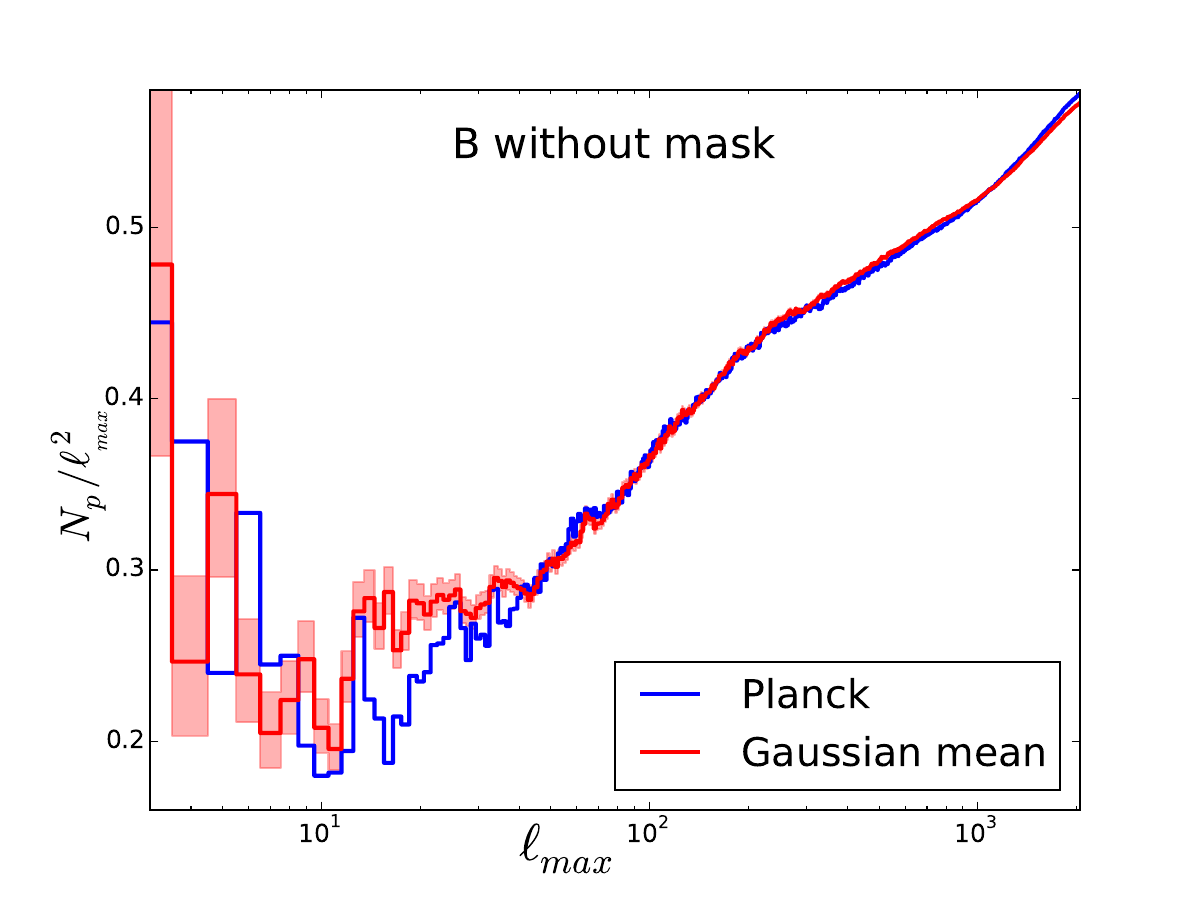}
  \includegraphics[width=0.49\textwidth]{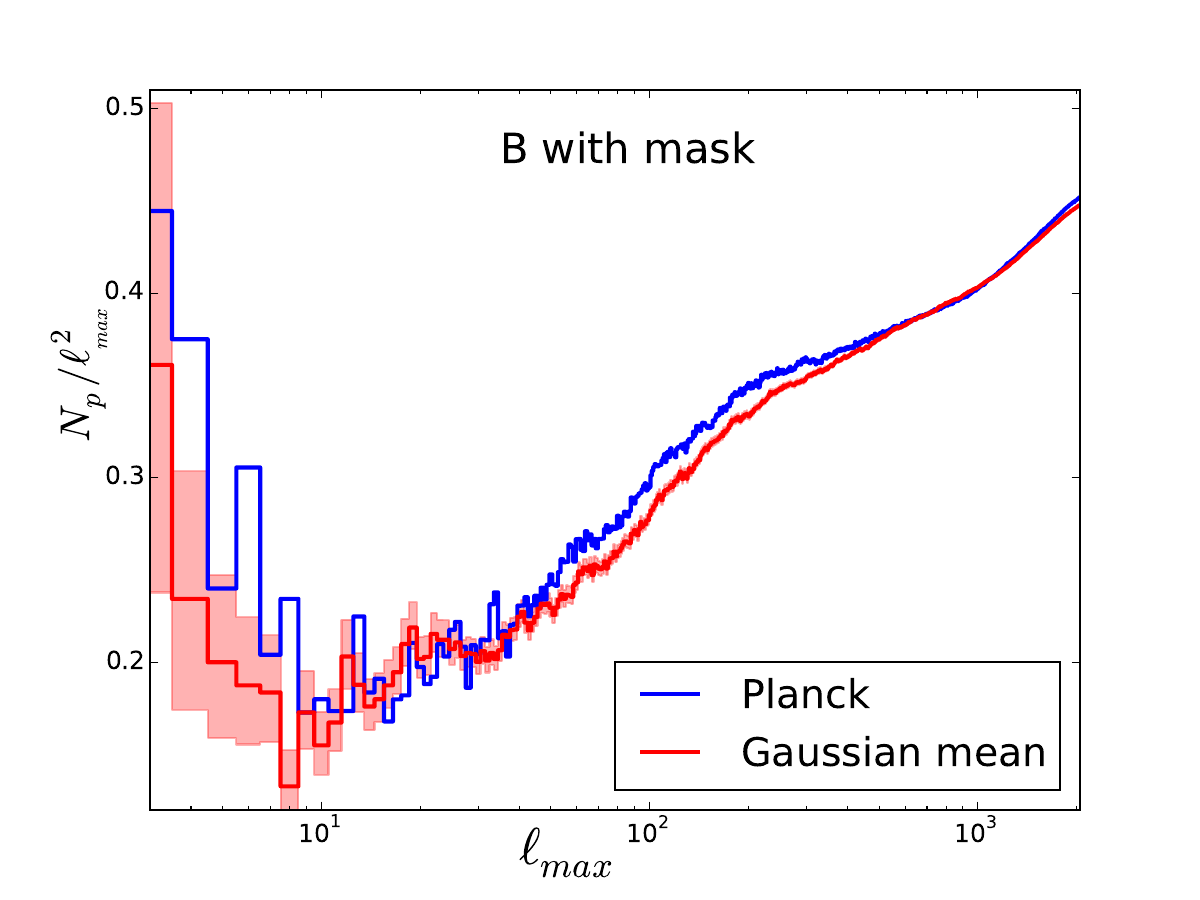}
  \includegraphics[width=0.49\textwidth]{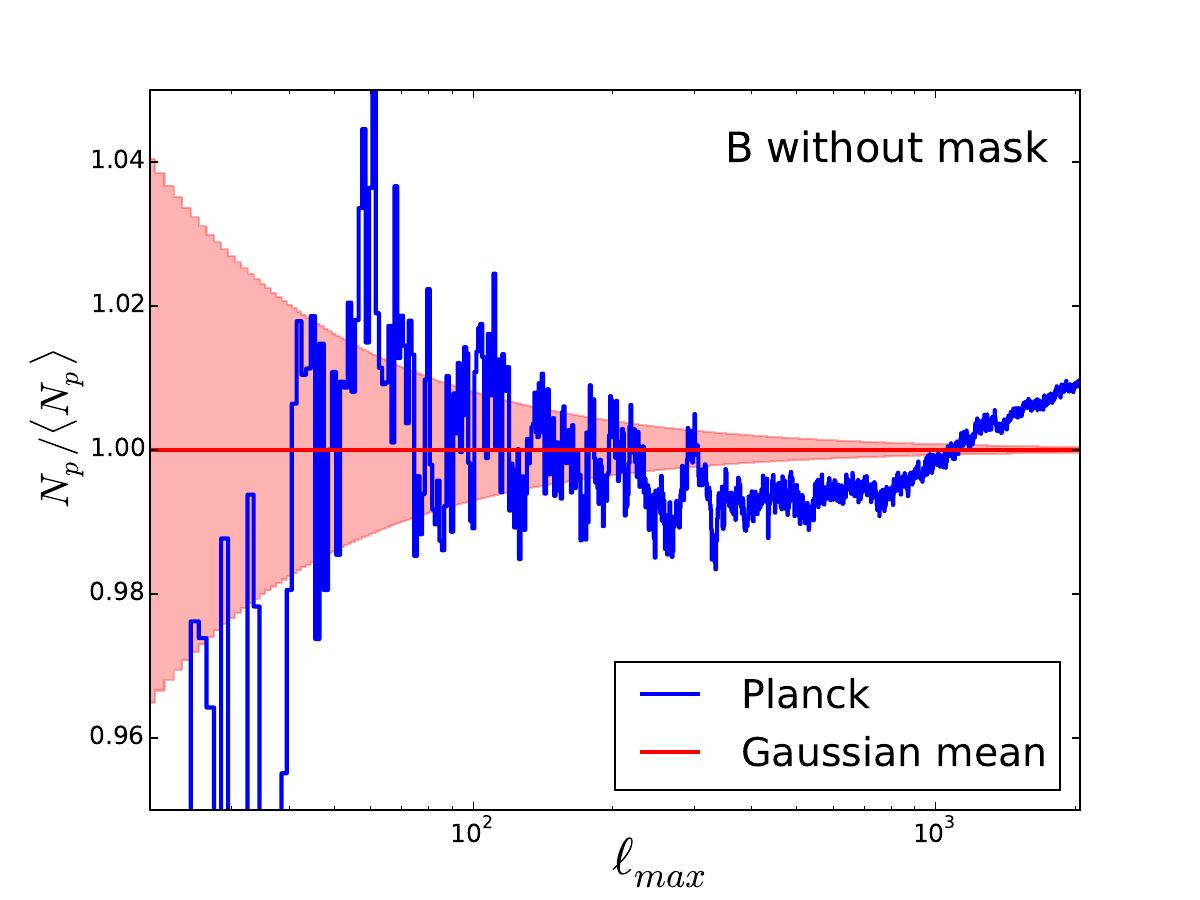}
  \includegraphics[width=0.49\textwidth]{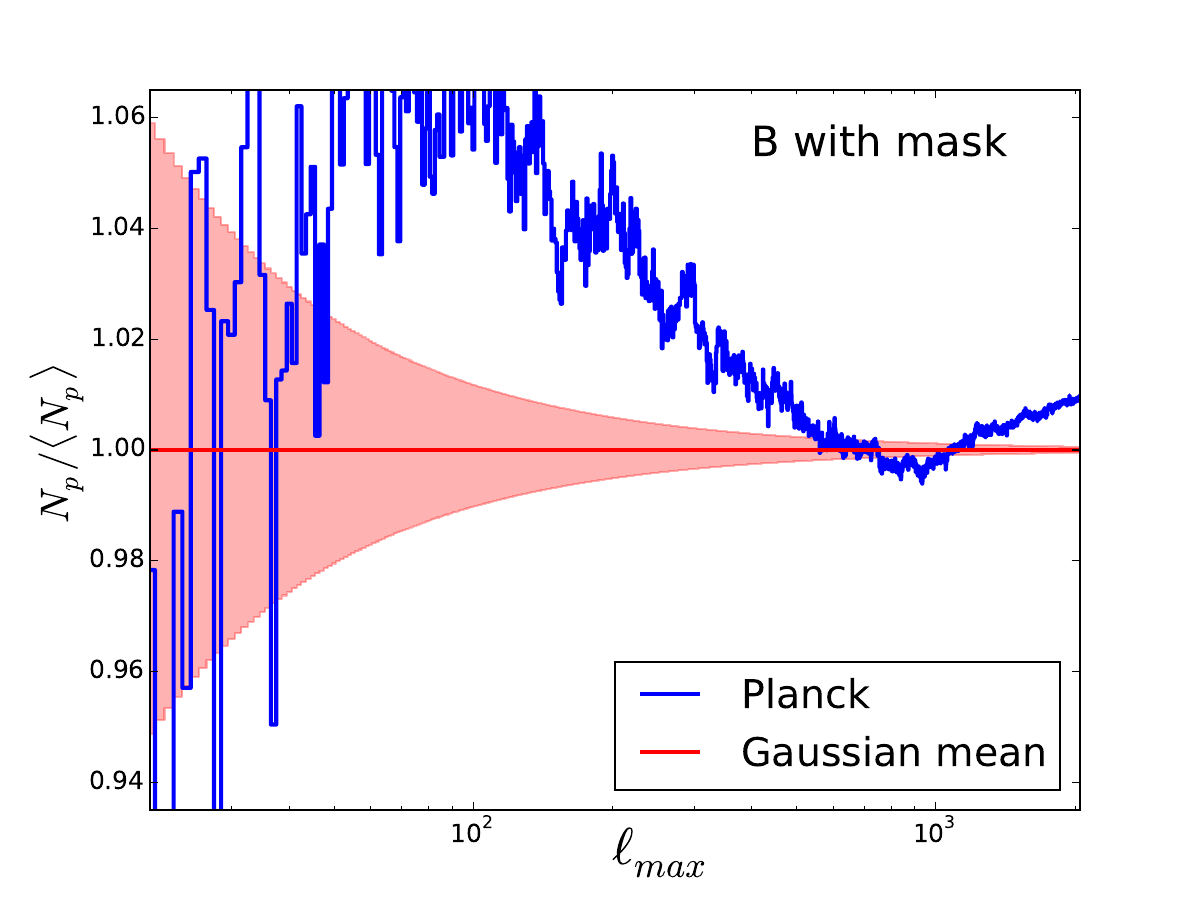}
  \caption{Gaussianity test for B polarization. The panels correspond to
    the same information as in Fig. 5, but for the B component.}
\end{figure*}
For the Planck data analysis we used the $Q$ and $U$ parameters from the
Planck Public Data Release 3 \citep{2020A&A...641A...4P}. In order to
achieve stability of our algorithm for finding
unpolarized points we had to use a map with equidistant pixelization in both
spherical coordinates. The number of pixels along $\varphi$,
as noted in Section II, must
at least 16 times exceed the number of harmonics $\ell_{max}$. Thus, to achieve
Planck resolution $\ell_{max}=2048$ we are dealing with a map
$(\theta,\varphi)$ of
$(16384\times 32768)$ pixels. Repixilization into our map from the
HEALpix format \citep{2005ApJ...622..759G} is fairly straightforward.
We used the $a_{\ell m}^{Q,U}$
coefficients to perform the spin-weighted spherical harmonic transform for
our map. Note that, similar to HEALpix, equidistant pixelization of
$\varphi_j$ makes it possible to apply the fast Fourier transform along
this coordinate, which significantly speeds up the process.

\begin{center}
{\it Data processing in the presence of the mask }
\end{center}

In our studies we used the polarization mask available in
\citep{2020A&A...641A...4P}.

For the Gaussianity test we need unpolarized points and a power spectrum.
Since Planck data is available for the entire sky,
we are able to completely separate the E and B polarization parts and
find such points separately for these two components both inside and
outside the mask.

When ignoring the mask, the $C_\ell^E$ and $C_\ell^B$ power spectra are
found directly by virtue of the orthogonality of the spin-weighted
spherical harmonics when integrated over the entire sky.

In case of using a mask, we have access to unpolarized points outside it.
For a correct analysis of Gaussianity, the power spectra
$\hat{C}_\ell^E$ and $\hat{C}_\ell^B$ must also be
found using only the unmasked data. The usual approach to estimate
the spectrum for just a part of the sky, is to use a two-point
correlation function. In this case, only unmasked pixels are used to
calculate correlations.  Expanding this function into
Legendre polynomials provides an estimate of the power spectrum for the
unmasked signal only. However, this approach does not completely eliminate
non-orthogonality and thus leads to not entirely correct values of
$\hat{C}_\ell^{E,B}$. This happens mainly because the angular distance
between different pixels does not exactly match the discrete values of
the correlation function argument.

To avoid this effect, we create maps for each $\ell$ separately 
(in particular for the scalar E):
\begin{equation}
  \begin{array}{l}
    E_\ell(\boldsymbol{\eta}_{i,j})=\sum\limits_{m=-\ell}^{\ell}a_{\ell m}^E
    \cdot Y_{\ell m}(\boldsymbol{\eta}_{i,j}),
    \end{array}
\end{equation}
where $a_{\ell m}^E$ are taken from Planck data and
$\boldsymbol{\eta}_{i,j}=(\theta_i,\varphi_j)$. This allows us to calculate
the power for each $\ell$ without the influence of other harmonics:
\begin{equation}
  \begin{array}{l}
    \vspace{0.3cm}
    \langle\left[E_\ell\right]^2\rangle=
    \frac{1}{S}\sum\limits_{i,j}
    \left[E_\ell(\boldsymbol{\eta}_{i,j})\right]^2
    h^2\sin\theta_i,\\
    \vspace{0.3cm}
    \langle[\hat{E}_\ell]^2\rangle=
    \frac{1}{\hat{S}}\sum\limits_{i,j}m(i,j)\left
     [E_\ell(\boldsymbol{\eta}_{i,j})\right]^2
     h^2\sin\theta_i,\\
     S=\sum\limits_{i,j}h^2\sin\theta_i,\hspace{0.3cm}
     \hat{S}=\sum\limits_{i,j}m(i,j)h^2\sin\theta_i,
    \end{array}
\end{equation}
\begin{figure*}[!htbp]
  \includegraphics[width=0.49\textwidth]{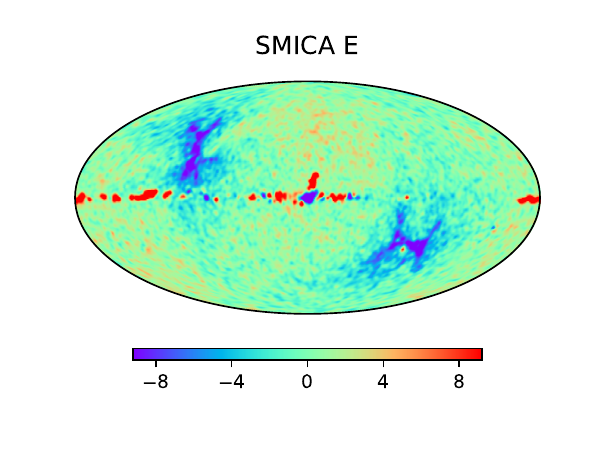}
  \includegraphics[width=0.49\textwidth]{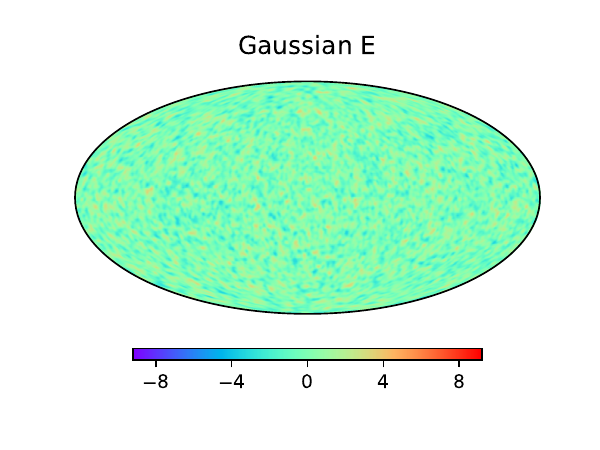}
  \includegraphics[width=0.49\textwidth]{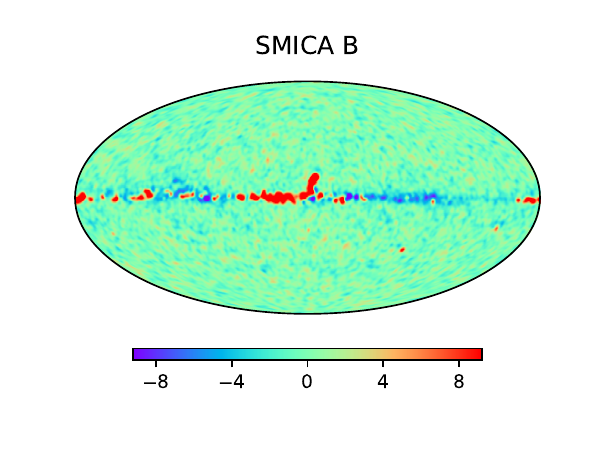}
  \includegraphics[width=0.49\textwidth]{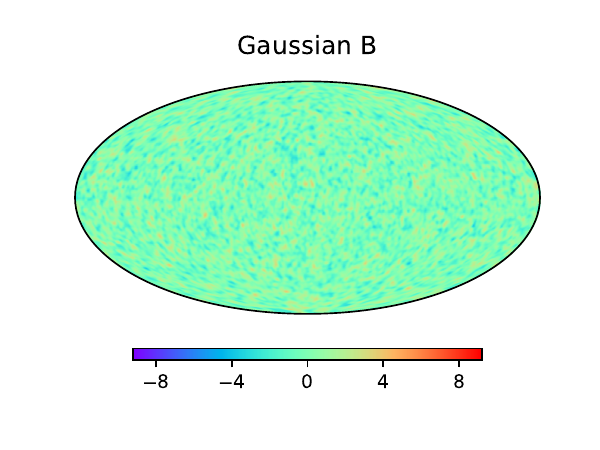}
  \caption{ Concentration maps of unpolarized points on the sky
    $\tilde{n}(\boldsymbol{\eta},\theta_0)$ with a smoothing angle
    $\theta=1^o$. {\it Upper left panel}: E polarization. The brightest red
    and blue spots highlight regions of non-Gaussianity with an excess or
    lack of null points, respectively. Due to the orientation of the
    telescope during scanning, the blue broad spots near the ecliptic poles
    contain less small-scale noise than the rest of the sky. In these spots,
    the signal is smoother and forms fewer unpolarized points.
    {\it Lower left panel}: B polarization. In the absence of any noticeable
    cosmological signal, the anisotropy of the noise does not affect the
    concentration of special points. {\it The upper right and lower right}
    panels present  $\tilde{n}(\boldsymbol{\eta},\theta_0)$  maps for the
    Gaussian
    E and B fields with spectra identical to those obtained by Planck.}
\end{figure*}
where the summation is performed over all pixels $(i,j)$ in the sky, $S$
is the area of the
entire sphere, $\hat{S}$ is the area of its unmasked part, the function
$m(i,j)=1$ outside the mask and equal to zero inside it. As a result, for
a correct Gaussianity test outside the mask, we get a corrected $\hat{C}_\ell^E$
spectrum:
\begin{equation}
  \begin{array}{l}
    \hat{C}_\ell^E=\frac{\langle[\hat{E}_\ell]^2\rangle}{
    \langle\left[E_\ell\right]^2\rangle}C_\ell^E.
    \end{array}
\end{equation}
The same should be done for spectrum $\hat{C}_\ell^B$.

\begin{center}
{\it General non-Gaussianity}
\end{center}

Similarly to the Earth's surface example, we tested the nature of the observed
signal for polarization maps at different angular scales using the
function $N_p(\ell_{max})$.

In order to generate Gaussian polarization maps with the Planck spectra
for E and B we used random Gaussian numbers $\tilde{a}_{\ell m}^{E,B}$
with zero mean and equal variances for a given $\ell$. The normalization of
these values for the full sky, i.e. without a mask, is as follows:
\begin{equation}
  \begin{array}{l}
    \sum\limits_{m=-\ell}^\ell
    \left(\tilde{a}_{\ell m}^{E,B}\right)\left(\tilde{a}_{\ell m}^{E,B}\right)^*
    =(2\ell+1)C_\ell^{E,B},
     \end{array}
\end{equation}
and the normalization in the presence of the polarization confidence mask
corresponds to the corrected spectra $\hat{C_\ell}^{E,B}$:
\begin{equation}
  \begin{array}{l}
    \sum\limits_{m=-\ell}^\ell
    \left(\tilde{a}_{\ell m}^{E,B}\right)\left(\tilde{a}_{\ell m}^{E,B}\right)^*
    =(2\ell+1)\hat{C}_\ell^{E,B}.
    \end{array}
\end{equation}

Using a large number of simulated maps for different values of $\ell_{max}$,
we found the mean number of unpolarized points for a Gaussian signal
$\langle N_p(\ell_{max})\rangle$, as well as the average deviations
above $\sigma^+(\ell_{max})$
and below $\sigma^-(\ell_{max})$ this value.
Note that $\sigma^+(\ell_{max})>\sigma^-(\ell_{max})$. This assymetry is not
surprising, since the upward deviations
can be arbitrarily large, while the downward deviations are limited because
$N_p$ cannot be negative. For large $\ell_{max}$ this distribution approaches
symmetry. The results of our calculations are shown in Figs. 4,5. It is easy
to see a quite strong non-Gaussianity for both signal components E and B
at almost all scales $\ell_{max}$. The reason for this is not only the presence
of unremoved foregrounds, but also, as will be shown below, the uneven
distribution of white noise on the Stokes parameters map. Besides, as one can
see, application of the mask does not improve the situation. The signal
remains strongly non-Gaussian.

\begin{center}
{\it Local non-Gaussianities}
\end{center}

A map of unpolarized points on the sky at a maximum Planck resolution
$\ell_{max}=2048$ can be represented as a two-dimensional function
$F(\boldsymbol{\eta})$:
\begin{equation}
  \begin{array}{l}
    \vspace{0.2cm}
    F(\boldsymbol{\eta})=
    \sum\limits_{p}\delta(\boldsymbol{\eta}-\boldsymbol{\eta}_p),\\
    \vspace{0.3cm}
    \delta(\boldsymbol{\eta})=1\hspace{0.3cm}if\hspace{0.5cm}
    \boldsymbol{\eta}=\boldsymbol{0},\\
    \delta(\boldsymbol{\eta})=0\hspace{0.3cm}if\hspace{0.5cm}
    \boldsymbol{\eta}\ne\boldsymbol{0},
    \end{array}
\end{equation}
where $\boldsymbol{\eta}_p$ denotes the location of the point with zero
polarization and the summation is performed over all points.
Thus, the discontinuous function $F$ at such points turns into unity and is
equal to zero at any other place. The total number of unpolarized points in the
entire sky exceeds 2,000,000 and they are scattered
almost uniformly. Thus, in order to detect possible anomalies in their
distribution, it is necessary to create a smooth continuous function of
the concentration of points $\tilde{n}$ by convolving
the map with a certain filter $\omega$ (This operation is similar to how a
continuous density distribution in space is created using the location of
galaxy clusters):
\begin{equation}
  \begin{array}{l}
    \vspace{0.3cm}
    \tilde{n}(\boldsymbol{\eta},\theta_0)=\frac{1}{4\pi}\int\limits_{4\pi}
    \omega(\boldsymbol{\eta'}\boldsymbol{\eta},\theta_0)\cdot
    F(\boldsymbol{\eta'})\cdot d\boldsymbol{\eta'},\\
    \omega(\boldsymbol{\eta'}\boldsymbol{\eta},\theta_0)=
    \frac{1}{\sqrt{2\pi}\theta_0}\exp\left[-\frac
     {acos(\boldsymbol{\eta'\eta})}{2\theta_0^{{}^2}}\right],
    \end{array}
\end{equation}
where $\boldsymbol{\eta\eta'}$ denotes the scalar product of vectors
$\boldsymbol{\eta}$ and $\boldsymbol{\eta'}$. The filter
$\omega(\boldsymbol{\eta\eta'},\theta_0)$ has a Gaussian
shape with half-width $\theta_0$. The integration with respect to
$d\boldsymbol{\eta'}=\sin\theta'd\theta' d\varphi'$ should be performed over the
entire sphere. The function $\tilde{n}(\boldsymbol{\eta},\theta_0)$ depends on
the smoothing scale $\theta_0$ and becomes equal to
$F(\boldsymbol{\eta})$ in Eq. (25) when $\theta_0\rightarrow 0$.

Fig. 6 shows the concentration maps $\tilde{n}(\boldsymbol{\eta},\theta_0)$
of Planck unpolarized points for the
E and B components at $\theta_0=1^o$ (left panel). The right panel shows
the example of the same maps for Gaussian E and B polarization, i.e. for
a random uniform distribution of phases. The spectra on the left and right
panels are identical and equal to the observed spectra $C_\ell^{E,B}$.

The E and B maps show obvious features of a clearly non-Gaussian nature
along the galaxy plane. In addition, the E mode map shows two zones near
the poles of Planck's sky scaning with a clear lack of unpolarized points.
These zones have a lower noise level than the rest of the sky and the signal
observed in them is smoother. This inevitably leads to a local increase in
the correlation scale and, consequently, to a decrease in the concentration
of zero polarization points, Eq. (13). Note that the concentration map of
special points exhibits almost the same shape of such regions as the error
map in the work \citep{2016A&A...594A..10P}. This shows that, among other things, this approach can
be used to estimate the signal-to-noise ratio for pixels on the map.

As for the B mode map, these regions do not appear there. This is explained
by the fact that the B component is dominated by small-scale noise, which
completely obscures the cosmological signal (if such a signal exists at all).

To simplify somewhat, we can say that the E mode consists of a cosmological
signal and a small-scale component. A local reduction in the amplitude of
this noise component changes the correlation scale. At the same time, if
the B mode contains only one noise component, then a local reduction in its
amplitude in certain regions does not change the correlation scale in them
and, as a consequence, does not affect the concentration of
unpolarized points.

\section{Conclusions}
Establishing the Gaussian or non-Gaussian nature of the initial perturbations
is one of the priority tasks of modern cosmology. Maps of CMB anisotropy and
polarization are the most convenient material for studying such fluctuations.
The approach for studying the statistics of non-polarized points on the sky
that we presented allows us to:\\
$\bullet$ Test for Gaussianity of the polarization maps at different angular
scales;\\
$\bullet$ Detect local non-Gaussianities caused by the presence of foregrounds
of various origins that have not been removed from the observational data;\\
$\bullet$ Independently estimate the signal-to-noise ratio in different
parts of the pixelized map.

The obvious non-Gaussianity of the E component of polarization that we
discovered means that more correct data processing is needed to clear the
polarization maps of foregrounds. Taking into account  the
absence of experiments with good sensitivity to date, we need more advanced
methods of filtering maps and careful ways of working with unevenly
distributed white noise.

It seems extremely interesting to study the statistics of unpolarized points
of different types (saddles, comets and beaks), which remains beyond the
scope of our article.

\newpage
The work is supported by the Project No. 36-2024 of LPI new scientific
groups.

\def\apj{Astrophys.~J}
\def\apjl{Astrophys.~J.,~Lett}
\def\apjs{Astrophys.~J.,~Supplement}
\def\an{Astron.~Nachr}     
\def\aap{Astron.~Astrophys}
\def\mnras{Mon.~Not.~R.~Astron.~Soc}
\def\pasp{Publ.~Astron.~Soc.~Pac}
\def\aaps{Astron.~and Astrophys.,~Suppl.~Ser}
\def\apss{Astrophys.~Space.~Sci}
\def\ibvs{Inf.~Bull.~Variable~Stars}
\def\japa{J.~Astrophys.~Astron}
\def\na{New~Astron}
\def\aspproc{Proc.~ASP~conf.~ser.}
\def\aspcs{ASP~Conf.~Ser}
\def\aj{Astron.~J}
\def\actaa{Acta Astron}
\def\araa{Ann.~Rev.~Astron.~Astrophys}
\def\caosp{Contrib.~Astron.~Obs.~Skalnat{\'e}~Pleso}
\def\pasj{Publ.~Astron.~Soc.~Jpn}
\def\memsai{Mem.~Soc.~Astron.~Ital}
\def\astl{Astron.~Letters}
\def\aipproc{Proc.~AIP~conf.~ser.}
\def\physrep{Physics Reports}
\def\jcap{Journal of Cosmology and Astroparticle Physics}
\def\baas{Bulletin of the AAS}
\def\ssr{Space~Sci.~Rev.}
\def\azh{Astronomicheskii Zhurnal}

\bibliography{a11pol.bib}

\providecommand{\noopsort}[1]{}\providecommand{\singleletter}[1]{#1}%
\begin{thebibliography}{53}%
\makeatletter
\providecommand \@ifxundefined [1]{%
 \@ifx{#1\undefined}
}%
\providecommand \@ifnum [1]{%
 \ifnum #1\expandafter \@firstoftwo
 \else \expandafter \@secondoftwo
 \fi
}%
\providecommand \@ifx [1]{%
 \ifx #1\expandafter \@firstoftwo
 \else \expandafter \@secondoftwo
 \fi
}%
\providecommand \natexlab [1]{#1}%
\providecommand \enquote  [1]{``#1''}%
\providecommand \bibnamefont  [1]{#1}%
\providecommand \bibfnamefont [1]{#1}%
\providecommand \citenamefont [1]{#1}%
\providecommand \href@noop [0]{\@secondoftwo}%
\providecommand \href [0]{\begingroup \@sanitize@url \@href}%
\providecommand \@href[1]{\@@startlink{#1}\@@href}%
\providecommand \@@href[1]{\endgroup#1\@@endlink}%
\providecommand \@sanitize@url [0]{\catcode `\\12\catcode `\$12\catcode
  `\&12\catcode `\#12\catcode `\^12\catcode `\_12\catcode `\%12\relax}%
\providecommand \@@startlink[1]{}%
\providecommand \@@endlink[0]{}%
\providecommand \url  [0]{\begingroup\@sanitize@url \@url }%
\providecommand \@url [1]{\endgroup\@href {#1}{\urlprefix }}%
\providecommand \urlprefix  [0]{URL }%
\providecommand \Eprint [0]{\href }%
\providecommand \doibase [0]{https://doi.org/}%
\providecommand \selectlanguage [0]{\@gobble}%
\providecommand \bibinfo  [0]{\@secondoftwo}%
\providecommand \bibfield  [0]{\@secondoftwo}%
\providecommand \translation [1]{[#1]}%
\providecommand \BibitemOpen [0]{}%
\providecommand \bibitemStop [0]{}%
\providecommand \bibitemNoStop [0]{.\EOS\space}%
\providecommand \EOS [0]{\spacefactor3000\relax}%
\providecommand \BibitemShut  [1]{\csname bibitem#1\endcsname}%
\let\auto@bib@innerbib\@empty
\bibitem [{\citenamefont {{Bennett}}\ \emph {et~al.}(2013)\citenamefont
  {{Bennett}}, \citenamefont {{Larson}}, \citenamefont {{Weiland}},
  \citenamefont {{Jarosik}}, \citenamefont {{Hinshaw}}, \citenamefont
  {{Odegard}}, \citenamefont {{Smith}}, \citenamefont {{Hill}}, \citenamefont
  {{Gold}}, \citenamefont {{Halpern}},\ and\ \citenamefont {{et
  al.}}}]{2013ApJS..208...20B}%
  \BibitemOpen
  \bibfield  {author} {\bibinfo {author} {\bibfnamefont {C.~L.}\ \bibnamefont
  {{Bennett}}}, \bibinfo {author} {\bibfnamefont {D.}~\bibnamefont {{Larson}}},
  \bibinfo {author} {\bibfnamefont {J.~L.}\ \bibnamefont {{Weiland}}}, \bibinfo
  {author} {\bibfnamefont {N.}~\bibnamefont {{Jarosik}}}, \bibinfo {author}
  {\bibfnamefont {G.}~\bibnamefont {{Hinshaw}}}, \bibinfo {author}
  {\bibfnamefont {N.}~\bibnamefont {{Odegard}}}, \bibinfo {author}
  {\bibfnamefont {K.~M.}\ \bibnamefont {{Smith}}}, \bibinfo {author}
  {\bibfnamefont {R.~S.}\ \bibnamefont {{Hill}}}, \bibinfo {author}
  {\bibfnamefont {B.}~\bibnamefont {{Gold}}}, \bibinfo {author} {\bibfnamefont
  {M.}~\bibnamefont {{Halpern}}},\ and\ \bibinfo {author} {\bibnamefont {{et
  al.}}},\ }\bibfield  {title} {\bibinfo {title} {{Nine-year Wilkinson
  Microwave Anisotropy Probe (WMAP) Observations: Final Maps and Results}},\
  }\href {https://doi.org/10.1088/0067-0049/208/2/20} {\bibfield  {journal}
  {\bibinfo  {journal} {\apjs}\ }\textbf {\bibinfo {volume} {208}},\ \bibinfo
  {eid} {20} (\bibinfo {year} {2013})},\ \Eprint
  {https://arxiv.org/abs/1212.5225} {arXiv:1212.5225 [astro-ph.CO]}
  \BibitemShut {NoStop}%
\bibitem [{\citenamefont {{Cheng}}\ \emph {et~al.}(2013)\citenamefont
  {{Cheng}}, \citenamefont {{Huang}},\ and\ \citenamefont
  {{Ma}}}]{2013JCAP...07..018C}%
  \BibitemOpen
  \bibfield  {author} {\bibinfo {author} {\bibfnamefont {C.}~\bibnamefont
  {{Cheng}}}, \bibinfo {author} {\bibfnamefont {Q.-G.}\ \bibnamefont
  {{Huang}}},\ and\ \bibinfo {author} {\bibfnamefont {Y.-Z.}\ \bibnamefont
  {{Ma}}},\ }\bibfield  {title} {\bibinfo {title} {{Constraints on single-field
  inflation with WMAP, SPT and ACT data {\textemdash} a last-minute stand
  before Planck}},\ }\href {https://doi.org/10.1088/1475-7516/2013/07/018}
  {\bibfield  {journal} {\bibinfo  {journal} {\jcap}\ }\textbf {\bibinfo
  {volume} {2013}},\ \bibinfo {eid} {018} (\bibinfo {year} {2013})},\ \Eprint
  {https://arxiv.org/abs/1303.4497} {arXiv:1303.4497 [astro-ph.CO]}
  \BibitemShut {NoStop}%
\bibitem [{\citenamefont {{Ade}}\ \emph {et~al.}(2014)\citenamefont {{Ade}},
  \citenamefont {{Aghanim}}, \citenamefont {{Armitage-Caplan}}, \citenamefont
  {{Arnaud}}, \citenamefont {{Ashdown}}, \citenamefont {{Atrio-Barandela}},
  \citenamefont {{Aumont}}, \citenamefont {{Baccigalupi}}, \citenamefont
  {{Banday}},\ and\ \citenamefont {et~al.
  (Planck~Collaboration)}}]{2014A&A...571A..15P}%
  \BibitemOpen
  \bibfield  {author} {\bibinfo {author} {\bibfnamefont {P.~A.~R.}\
  \bibnamefont {{Ade}}}, \bibinfo {author} {\bibfnamefont {N.}~\bibnamefont
  {{Aghanim}}}, \bibinfo {author} {\bibfnamefont {C.}~\bibnamefont
  {{Armitage-Caplan}}}, \bibinfo {author} {\bibfnamefont {M.}~\bibnamefont
  {{Arnaud}}}, \bibinfo {author} {\bibfnamefont {M.}~\bibnamefont {{Ashdown}}},
  \bibinfo {author} {\bibfnamefont {F.}~\bibnamefont {{Atrio-Barandela}}},
  \bibinfo {author} {\bibfnamefont {J.}~\bibnamefont {{Aumont}}}, \bibinfo
  {author} {\bibfnamefont {C.}~\bibnamefont {{Baccigalupi}}}, \bibinfo {author}
  {\bibfnamefont {A.~J.}\ \bibnamefont {{Banday}}},\ and\ \bibinfo {author}
  {\bibnamefont {et~al. (Planck~Collaboration)}},\ }\bibfield  {title}
  {\bibinfo {title} {{Planck 2013 results. XV. CMB power spectra and
  likelihood}},\ }\href {https://doi.org/10.1051/0004-6361/201321573}
  {\bibfield  {journal} {\bibinfo  {journal} {\aap}\ }\textbf {\bibinfo
  {volume} {571}},\ \bibinfo {eid} {A15} (\bibinfo {year} {2014})},\ \Eprint
  {https://arxiv.org/abs/1303.5075} {arXiv:1303.5075} \BibitemShut {NoStop}%
\bibitem [{\citenamefont {{Planck Collaboration}}\ \emph
  {et~al.}(2020{\natexlab{a}})\citenamefont {{Planck Collaboration}},
  \citenamefont {{Akrami}}, \citenamefont {{Ashdown}}, \citenamefont
  {{Aumont}}, \citenamefont {{Baccigalupi}}, \citenamefont {{Ballardini}},
  \citenamefont {{Banday}}, \citenamefont {{Barreiro}}, \citenamefont
  {{Bartolo}}, \citenamefont {{Basak}},\ and\ \citenamefont
  {et~al.}}]{2018arXiv180706208P}%
  \BibitemOpen
  \bibfield  {author} {\bibinfo {author} {\bibnamefont {{Planck
  Collaboration}}}, \bibinfo {author} {\bibfnamefont {Y.}~\bibnamefont
  {{Akrami}}}, \bibinfo {author} {\bibfnamefont {M.}~\bibnamefont {{Ashdown}}},
  \bibinfo {author} {\bibfnamefont {J.}~\bibnamefont {{Aumont}}}, \bibinfo
  {author} {\bibfnamefont {C.}~\bibnamefont {{Baccigalupi}}}, \bibinfo {author}
  {\bibfnamefont {M.}~\bibnamefont {{Ballardini}}}, \bibinfo {author}
  {\bibfnamefont {A.~J.}\ \bibnamefont {{Banday}}}, \bibinfo {author}
  {\bibfnamefont {R.~B.}\ \bibnamefont {{Barreiro}}}, \bibinfo {author}
  {\bibfnamefont {N.}~\bibnamefont {{Bartolo}}}, \bibinfo {author}
  {\bibfnamefont {S.}~\bibnamefont {{Basak}}},\ and\ \bibinfo {author}
  {\bibnamefont {et~al.}},\ }\bibfield  {title} {\bibinfo {title} {{Planck 2018
  results. IV. Diffuse component separation}},\ }\href
  {https://doi.org/10.1051/0004-6361/201833881} {\bibfield  {journal} {\bibinfo
   {journal} {\aap}\ }\textbf {\bibinfo {volume} {641}},\ \bibinfo {eid} {A4}
  (\bibinfo {year} {2020}{\natexlab{a}})},\ \Eprint
  {https://arxiv.org/abs/1807.06208} {arXiv:1807.06208 [astro-ph.CO]}
  \BibitemShut {NoStop}%
\bibitem [{\citenamefont {{Planck Collaboration}}\ \emph
  {et~al.}(2020{\natexlab{b}})\citenamefont {{Planck Collaboration}},
  \citenamefont {{Akrami}}, \citenamefont {{Ashdown}}, \citenamefont
  {{Aumont}}, \citenamefont {{Baccigalupi}}, \citenamefont {{Ballardini}},
  \citenamefont {{Banday}}, \citenamefont {{Barreiro}}, \citenamefont
  {{Bartolo}}, \citenamefont {{Basak}},\ and\ \citenamefont
  {et~al.}}]{2020A&A...641A...7P}%
  \BibitemOpen
  \bibfield  {author} {\bibinfo {author} {\bibnamefont {{Planck
  Collaboration}}}, \bibinfo {author} {\bibfnamefont {Y.}~\bibnamefont
  {{Akrami}}}, \bibinfo {author} {\bibfnamefont {M.}~\bibnamefont {{Ashdown}}},
  \bibinfo {author} {\bibfnamefont {J.}~\bibnamefont {{Aumont}}}, \bibinfo
  {author} {\bibfnamefont {C.}~\bibnamefont {{Baccigalupi}}}, \bibinfo {author}
  {\bibfnamefont {M.}~\bibnamefont {{Ballardini}}}, \bibinfo {author}
  {\bibfnamefont {A.~J.}\ \bibnamefont {{Banday}}}, \bibinfo {author}
  {\bibfnamefont {R.~B.}\ \bibnamefont {{Barreiro}}}, \bibinfo {author}
  {\bibfnamefont {N.}~\bibnamefont {{Bartolo}}}, \bibinfo {author}
  {\bibfnamefont {S.}~\bibnamefont {{Basak}}},\ and\ \bibinfo {author}
  {\bibnamefont {et~al.}},\ }\bibfield  {title} {\bibinfo {title} {{Planck 2018
  results. VII. Isotropy and statistics of the CMB}},\ }\href
  {https://doi.org/10.1051/0004-6361/201935201} {\bibfield  {journal} {\bibinfo
   {journal} {\aap}\ }\textbf {\bibinfo {volume} {641}},\ \bibinfo {eid} {A7}
  (\bibinfo {year} {2020}{\natexlab{b}})},\ \Eprint
  {https://arxiv.org/abs/1906.02552} {arXiv:1906.02552 [astro-ph.CO]}
  \BibitemShut {NoStop}%
\bibitem [{\citenamefont {{Linde}}\ and\ \citenamefont
  {{Mukhanov}}(1997)}]{1997PhRvD..56..535L}%
  \BibitemOpen
  \bibfield  {author} {\bibinfo {author} {\bibfnamefont {A.}~\bibnamefont
  {{Linde}}}\ and\ \bibinfo {author} {\bibfnamefont {V.}~\bibnamefont
  {{Mukhanov}}},\ }\bibfield  {title} {\bibinfo {title} {{Non-Gaussian
  isocurvature perturbations from inflation}},\ }\href
  {https://doi.org/10.1103/PhysRevD.56.R535} {\bibfield  {journal} {\bibinfo
  {journal} {\prd}\ }\textbf {\bibinfo {volume} {56}},\ \bibinfo {pages} {R535}
  (\bibinfo {year} {1997})},\ \Eprint {https://arxiv.org/abs/astro-ph/9610219}
  {arXiv:astro-ph/9610219 [astro-ph]} \BibitemShut {NoStop}%
\bibitem [{\citenamefont {{Ortolan}}\ \emph {et~al.}(1988)\citenamefont
  {{Ortolan}}, \citenamefont {{Lucchin}},\ and\ \citenamefont
  {{Matarrese}}}]{1988PhRvD..38..465O}%
  \BibitemOpen
  \bibfield  {author} {\bibinfo {author} {\bibfnamefont {A.}~\bibnamefont
  {{Ortolan}}}, \bibinfo {author} {\bibfnamefont {F.}~\bibnamefont
  {{Lucchin}}},\ and\ \bibinfo {author} {\bibfnamefont {S.}~\bibnamefont
  {{Matarrese}}},\ }\bibfield  {title} {\bibinfo {title} {{Non-Gaussian
  perturbations from inflationary dynamics}},\ }\href
  {https://doi.org/10.1103/PhysRevD.38.465} {\bibfield  {journal} {\bibinfo
  {journal} {\prd}\ }\textbf {\bibinfo {volume} {38}},\ \bibinfo {pages} {465}
  (\bibinfo {year} {1988})}\BibitemShut {NoStop}%
\bibitem [{\citenamefont {{Doroshkevich}}(1970)}]{1970Ap......6..320D}%
  \BibitemOpen
  \bibfield  {author} {\bibinfo {author} {\bibfnamefont {A.~G.}\ \bibnamefont
  {{Doroshkevich}}},\ }\bibfield  {title} {\bibinfo {title} {{Spatial structure
  of perturbations and origin of galactic rotation in fluctuation theory}},\
  }\href {https://doi.org/10.1007/BF01001625} {\bibfield  {journal} {\bibinfo
  {journal} {Astrophysics}\ }\textbf {\bibinfo {volume} {6}},\ \bibinfo {pages}
  {320} (\bibinfo {year} {1970})}\BibitemShut {NoStop}%
\bibitem [{\citenamefont {{Bardeen}}\ \emph {et~al.}(1986)\citenamefont
  {{Bardeen}}, \citenamefont {{Bond}}, \citenamefont {{Kaiser}},\ and\
  \citenamefont {{Szalay}}}]{1986ApJ...304...15B}%
  \BibitemOpen
  \bibfield  {author} {\bibinfo {author} {\bibfnamefont {J.~M.}\ \bibnamefont
  {{Bardeen}}}, \bibinfo {author} {\bibfnamefont {J.~R.}\ \bibnamefont
  {{Bond}}}, \bibinfo {author} {\bibfnamefont {N.}~\bibnamefont {{Kaiser}}},\
  and\ \bibinfo {author} {\bibfnamefont {A.~S.}\ \bibnamefont {{Szalay}}},\
  }\bibfield  {title} {\bibinfo {title} {{The Statistics of Peaks of Gaussian
  Random Fields}},\ }\href {https://doi.org/10.1086/164143} {\bibfield
  {journal} {\bibinfo  {journal} {\apj}\ }\textbf {\bibinfo {volume} {304}},\
  \bibinfo {pages} {15} (\bibinfo {year} {1986})}\BibitemShut {NoStop}%
\bibitem [{\citenamefont {{Bond}}\ and\ \citenamefont
  {{Efstathiou}}(1987)}]{1987MNRAS.226..655B}%
  \BibitemOpen
  \bibfield  {author} {\bibinfo {author} {\bibfnamefont {J.~R.}\ \bibnamefont
  {{Bond}}}\ and\ \bibinfo {author} {\bibfnamefont {G.}~\bibnamefont
  {{Efstathiou}}},\ }\bibfield  {title} {\bibinfo {title} {{The statistics of
  cosmic background radiation fluctuations}},\ }\href
  {https://doi.org/10.1093/mnras/226.3.655} {\bibfield  {journal} {\bibinfo
  {journal} {\mnras}\ }\textbf {\bibinfo {volume} {226}},\ \bibinfo {pages}
  {655} (\bibinfo {year} {1987})}\BibitemShut {NoStop}%
\bibitem [{\citenamefont {{Gott}}\ \emph {et~al.}(1990)\citenamefont {{Gott}},
  \citenamefont {{Park}}, \citenamefont {{Juszkiewicz}}, \citenamefont
  {{Bies}}, \citenamefont {{Bennett}}, \citenamefont {{Bouchet}},\ and\
  \citenamefont {{Stebbins}}}]{1990ApJ...352....1G}%
  \BibitemOpen
  \bibfield  {author} {\bibinfo {author} {\bibfnamefont {I.}~\bibnamefont
  {{Gott}}, \bibfnamefont {J.~Richard}}, \bibinfo {author} {\bibfnamefont
  {C.}~\bibnamefont {{Park}}}, \bibinfo {author} {\bibfnamefont
  {R.}~\bibnamefont {{Juszkiewicz}}}, \bibinfo {author} {\bibfnamefont {W.~E.}\
  \bibnamefont {{Bies}}}, \bibinfo {author} {\bibfnamefont {D.~P.}\
  \bibnamefont {{Bennett}}}, \bibinfo {author} {\bibfnamefont {F.~R.}\
  \bibnamefont {{Bouchet}}},\ and\ \bibinfo {author} {\bibfnamefont
  {A.}~\bibnamefont {{Stebbins}}},\ }\bibfield  {title} {\bibinfo {title}
  {{Topology of Microwave Background Fluctuations: Theory}},\ }\href
  {https://doi.org/10.1086/168511} {\bibfield  {journal} {\bibinfo  {journal}
  {\apj}\ }\textbf {\bibinfo {volume} {352}},\ \bibinfo {pages} {1} (\bibinfo
  {year} {1990})}\BibitemShut {NoStop}%
\bibitem [{\citenamefont {{Schmalzing}}\ and\ \citenamefont
  {{Gorski}}(1998)}]{1998MNRAS.297..355S}%
  \BibitemOpen
  \bibfield  {author} {\bibinfo {author} {\bibfnamefont {J.}~\bibnamefont
  {{Schmalzing}}}\ and\ \bibinfo {author} {\bibfnamefont {K.~M.}\ \bibnamefont
  {{Gorski}}},\ }\bibfield  {title} {\bibinfo {title} {{Minkowski functionals
  used in the morphological analysis of cosmic microwave background anisotropy
  maps}},\ }\href {https://doi.org/10.1046/j.1365-8711.1998.01467.x} {\bibfield
   {journal} {\bibinfo  {journal} {\mnras}\ }\textbf {\bibinfo {volume}
  {297}},\ \bibinfo {pages} {355} (\bibinfo {year} {1998})},\ \Eprint
  {https://arxiv.org/abs/astro-ph/9710185} {arXiv:astro-ph/9710185 [astro-ph]}
  \BibitemShut {NoStop}%
\bibitem [{\citenamefont {{Naselsky}}\ and\ \citenamefont
  {{Novikov}}(1998)}]{1998ApJ...507...31N}%
  \BibitemOpen
  \bibfield  {author} {\bibinfo {author} {\bibfnamefont {P.~D.}\ \bibnamefont
  {{Naselsky}}}\ and\ \bibinfo {author} {\bibfnamefont {D.~I.}\ \bibnamefont
  {{Novikov}}},\ }\bibfield  {title} {\bibinfo {title} {{General Statistical
  Properties of the Cosmic Microwave Background Polarization Field}},\ }\href
  {https://doi.org/10.1086/306294} {\bibfield  {journal} {\bibinfo  {journal}
  {\apj}\ }\textbf {\bibinfo {volume} {507}},\ \bibinfo {pages} {31} (\bibinfo
  {year} {1998})},\ \Eprint {https://arxiv.org/abs/astro-ph/9801285}
  {arXiv:astro-ph/9801285 [astro-ph]} \BibitemShut {NoStop}%
\bibitem [{\citenamefont {{Zhao}}(2014)}]{2014RAA....14..625Z}%
  \BibitemOpen
  \bibfield  {author} {\bibinfo {author} {\bibfnamefont {W.}~\bibnamefont
  {{Zhao}}},\ }\bibfield  {title} {\bibinfo {title} {{Probing the CMB cold spot
  through local Minkowski functionals}},\ }\href
  {https://doi.org/10.1088/1674-4527/14/6/002} {\bibfield  {journal} {\bibinfo
  {journal} {Research in Astronomy and Astrophysics}\ }\textbf {\bibinfo
  {volume} {14}},\ \bibinfo {eid} {625-634} (\bibinfo {year} {2014})},\ \Eprint
  {https://arxiv.org/abs/1209.4021} {arXiv:1209.4021 [astro-ph.CO]}
  \BibitemShut {NoStop}%
\bibitem [{\citenamefont {{Ganesan}}\ \emph {et~al.}(2015)\citenamefont
  {{Ganesan}}, \citenamefont {{Chingangbam}}, \citenamefont {{Yogendran}},\
  and\ \citenamefont {{Park}}}]{2015JCAP...02..028G}%
  \BibitemOpen
  \bibfield  {author} {\bibinfo {author} {\bibfnamefont {V.}~\bibnamefont
  {{Ganesan}}}, \bibinfo {author} {\bibfnamefont {P.}~\bibnamefont
  {{Chingangbam}}}, \bibinfo {author} {\bibfnamefont {K.~P.}\ \bibnamefont
  {{Yogendran}}},\ and\ \bibinfo {author} {\bibfnamefont {C.}~\bibnamefont
  {{Park}}},\ }\bibfield  {title} {\bibinfo {title} {{Primordial non-Gaussian
  signatures in CMB polarization}},\ }\href
  {https://doi.org/10.1088/1475-7516/2015/02/028} {\bibfield  {journal}
  {\bibinfo  {journal} {\jcap}\ }\textbf {\bibinfo {volume} {2015}},\ \bibinfo
  {pages} {028} (\bibinfo {year} {2015})},\ \Eprint
  {https://arxiv.org/abs/1411.5256} {arXiv:1411.5256 [astro-ph.CO]}
  \BibitemShut {NoStop}%
\bibitem [{\citenamefont {{Santos}}\ \emph {et~al.}(2016)\citenamefont
  {{Santos}}, \citenamefont {{Wang}},\ and\ \citenamefont
  {{Zhao}}}]{2016JCAP...07..029S}%
  \BibitemOpen
  \bibfield  {author} {\bibinfo {author} {\bibfnamefont {L.}~\bibnamefont
  {{Santos}}}, \bibinfo {author} {\bibfnamefont {K.}~\bibnamefont {{Wang}}},\
  and\ \bibinfo {author} {\bibfnamefont {W.}~\bibnamefont {{Zhao}}},\
  }\bibfield  {title} {\bibinfo {title} {{Probing the statistical properties of
  CMB B-mode polarization through Minkowski functionals}},\ }\href
  {https://doi.org/10.1088/1475-7516/2016/07/029} {\bibfield  {journal}
  {\bibinfo  {journal} {\jcap}\ }\textbf {\bibinfo {volume} {2016}},\ \bibinfo
  {eid} {029} (\bibinfo {year} {2016})},\ \Eprint
  {https://arxiv.org/abs/1510.07779} {arXiv:1510.07779 [astro-ph.CO]}
  \BibitemShut {NoStop}%
\bibitem [{\citenamefont {{Chingangbam}}\ \emph {et~al.}(2017)\citenamefont
  {{Chingangbam}}, \citenamefont {{Ganesan}}, \citenamefont {{Yogendran}},\
  and\ \citenamefont {{Park}}}]{2017PhLB..771...67C}%
  \BibitemOpen
  \bibfield  {author} {\bibinfo {author} {\bibfnamefont {P.}~\bibnamefont
  {{Chingangbam}}}, \bibinfo {author} {\bibfnamefont {V.}~\bibnamefont
  {{Ganesan}}}, \bibinfo {author} {\bibfnamefont {K.~P.}\ \bibnamefont
  {{Yogendran}}},\ and\ \bibinfo {author} {\bibfnamefont {C.}~\bibnamefont
  {{Park}}},\ }\bibfield  {title} {\bibinfo {title} {{On Minkowski Functionals
  of CMB polarization}},\ }\href
  {https://doi.org/10.1016/j.physletb.2017.05.030} {\bibfield  {journal}
  {\bibinfo  {journal} {Physics Letters B}\ }\textbf {\bibinfo {volume}
  {771}},\ \bibinfo {pages} {67} (\bibinfo {year} {2017})},\ \Eprint
  {https://arxiv.org/abs/1705.04454} {arXiv:1705.04454 [astro-ph.CO]}
  \BibitemShut {NoStop}%
\bibitem [{\citenamefont {{Joby}}\ \emph {et~al.}(2019)\citenamefont {{Joby}},
  \citenamefont {{Chingangbam}}, \citenamefont {{Ghosh}}, \citenamefont
  {{Ganesan}},\ and\ \citenamefont {{Ravikumar}}}]{2019JCAP...01..009J}%
  \BibitemOpen
  \bibfield  {author} {\bibinfo {author} {\bibfnamefont {P.~K.}\ \bibnamefont
  {{Joby}}}, \bibinfo {author} {\bibfnamefont {P.}~\bibnamefont
  {{Chingangbam}}}, \bibinfo {author} {\bibfnamefont {T.}~\bibnamefont
  {{Ghosh}}}, \bibinfo {author} {\bibfnamefont {V.}~\bibnamefont {{Ganesan}}},\
  and\ \bibinfo {author} {\bibfnamefont {C.~D.}\ \bibnamefont {{Ravikumar}}},\
  }\bibfield  {title} {\bibinfo {title} {{Search for anomalous alignments of
  structures in Planck data using Minkowski Tensors}},\ }\href
  {https://doi.org/10.1088/1475-7516/2019/01/009} {\bibfield  {journal}
  {\bibinfo  {journal} {\jcap}\ }\textbf {\bibinfo {volume} {2019}},\ \bibinfo
  {eid} {009} (\bibinfo {year} {2019})},\ \Eprint
  {https://arxiv.org/abs/1807.01306} {arXiv:1807.01306 [astro-ph.CO]}
  \BibitemShut {NoStop}%
\bibitem [{\citenamefont {{Kochappan}}\ \emph {et~al.}(2021)\citenamefont
  {{Kochappan}}, \citenamefont {{Sen}}, \citenamefont {{Ghosh}}, \citenamefont
  {{Chingangbam}},\ and\ \citenamefont {{Basak}}}]{2021PhRvD.103l3523K}%
  \BibitemOpen
  \bibfield  {author} {\bibinfo {author} {\bibfnamefont {J.~P.}\ \bibnamefont
  {{Kochappan}}}, \bibinfo {author} {\bibfnamefont {A.}~\bibnamefont {{Sen}}},
  \bibinfo {author} {\bibfnamefont {T.}~\bibnamefont {{Ghosh}}}, \bibinfo
  {author} {\bibfnamefont {P.}~\bibnamefont {{Chingangbam}}},\ and\ \bibinfo
  {author} {\bibfnamefont {S.}~\bibnamefont {{Basak}}},\ }\bibfield  {title}
  {\bibinfo {title} {{Application of the contour Minkowski tensor and D
  statistic to the Planck E -mode data}},\ }\href
  {https://doi.org/10.1103/PhysRevD.103.123523} {\bibfield  {journal} {\bibinfo
   {journal} {\prd}\ }\textbf {\bibinfo {volume} {103}},\ \bibinfo {eid}
  {123523} (\bibinfo {year} {2021})},\ \Eprint
  {https://arxiv.org/abs/2106.05757} {arXiv:2106.05757 [astro-ph.CO]}
  \BibitemShut {NoStop}%
\bibitem [{\citenamefont {{Carones}}\ \emph {et~al.}(2024)\citenamefont
  {{Carones}}, \citenamefont {{Carr{\'o}nDuque}}, \citenamefont {{Marinucci}},
  \citenamefont {{Migliaccio}},\ and\ \citenamefont
  {{Vittorio}}}]{2024MNRAS.527..756C}%
  \BibitemOpen
  \bibfield  {author} {\bibinfo {author} {\bibfnamefont {A.}~\bibnamefont
  {{Carones}}}, \bibinfo {author} {\bibfnamefont {J.}~\bibnamefont
  {{Carr{\'o}nDuque}}}, \bibinfo {author} {\bibfnamefont {D.}~\bibnamefont
  {{Marinucci}}}, \bibinfo {author} {\bibfnamefont {M.}~\bibnamefont
  {{Migliaccio}}},\ and\ \bibinfo {author} {\bibfnamefont {N.}~\bibnamefont
  {{Vittorio}}},\ }\bibfield  {title} {\bibinfo {title} {{Minkowski functionals
  of CMB polarization intensity with PYNKOWSKI: theory and application to
  Planck and future data}},\ }\href {https://doi.org/10.1093/mnras/stad3002}
  {\bibfield  {journal} {\bibinfo  {journal} {\mnras}\ }\textbf {\bibinfo
  {volume} {527}},\ \bibinfo {pages} {756} (\bibinfo {year} {2024})},\ \Eprint
  {https://arxiv.org/abs/2211.07562} {arXiv:2211.07562 [astro-ph.CO]}
  \BibitemShut {NoStop}%
\bibitem [{\citenamefont {{Luo}}\ and\ \citenamefont
  {{Schramm}}(1993{\natexlab{a}})}]{1993PhRvL..71.1124L}%
  \BibitemOpen
  \bibfield  {author} {\bibinfo {author} {\bibfnamefont {X.}~\bibnamefont
  {{Luo}}}\ and\ \bibinfo {author} {\bibfnamefont {D.~N.}\ \bibnamefont
  {{Schramm}}},\ }\bibfield  {title} {\bibinfo {title} {{Testing for the
  Gaussian nature of cosmological density perturbations through the three-point
  temperature correlation function}},\ }\href
  {https://doi.org/10.1103/PhysRevLett.71.1124} {\bibfield  {journal} {\bibinfo
   {journal} {\prl}\ }\textbf {\bibinfo {volume} {71}},\ \bibinfo {pages}
  {1124} (\bibinfo {year} {1993}{\natexlab{a}})},\ \Eprint
  {https://arxiv.org/abs/astro-ph/9305009} {arXiv:astro-ph/9305009 [astro-ph]}
  \BibitemShut {NoStop}%
\bibitem [{\citenamefont {{Luo}}(1994)}]{1994ApJ...427L..71L}%
  \BibitemOpen
  \bibfield  {author} {\bibinfo {author} {\bibfnamefont {X.}~\bibnamefont
  {{Luo}}},\ }\bibfield  {title} {\bibinfo {title} {{The Angular Bispectrum of
  the Cosmic Microwave Background}},\ }\href {https://doi.org/10.1086/187367}
  {\bibfield  {journal} {\bibinfo  {journal} {\apjl}\ }\textbf {\bibinfo
  {volume} {427}},\ \bibinfo {pages} {L71} (\bibinfo {year} {1994})},\ \Eprint
  {https://arxiv.org/abs/astro-ph/9312004} {arXiv:astro-ph/9312004 [astro-ph]}
  \BibitemShut {NoStop}%
\bibitem [{\citenamefont {{Planck Collaboration}}\ \emph
  {et~al.}(2020{\natexlab{c}})\citenamefont {{Planck Collaboration}},
  \citenamefont {{Akrami}}, \citenamefont {{Arroja}}, \citenamefont
  {{Ashdown}}, \citenamefont {{Aumont}}, \citenamefont {{Baccigalupi}},
  \citenamefont {{Ballardini}}, \citenamefont {{Banday}}, \citenamefont
  {{Barreiro}}, \citenamefont {{Bartolo}},\ and\ \citenamefont
  {et~al.}}]{2020A&A...641A...9P}%
  \BibitemOpen
  \bibfield  {author} {\bibinfo {author} {\bibnamefont {{Planck
  Collaboration}}}, \bibinfo {author} {\bibfnamefont {Y.}~\bibnamefont
  {{Akrami}}}, \bibinfo {author} {\bibfnamefont {F.}~\bibnamefont {{Arroja}}},
  \bibinfo {author} {\bibfnamefont {M.}~\bibnamefont {{Ashdown}}}, \bibinfo
  {author} {\bibfnamefont {J.}~\bibnamefont {{Aumont}}}, \bibinfo {author}
  {\bibfnamefont {C.}~\bibnamefont {{Baccigalupi}}}, \bibinfo {author}
  {\bibfnamefont {M.}~\bibnamefont {{Ballardini}}}, \bibinfo {author}
  {\bibfnamefont {A.~J.}\ \bibnamefont {{Banday}}}, \bibinfo {author}
  {\bibfnamefont {R.~B.}\ \bibnamefont {{Barreiro}}}, \bibinfo {author}
  {\bibfnamefont {N.}~\bibnamefont {{Bartolo}}},\ and\ \bibinfo {author}
  {\bibnamefont {et~al.}},\ }\bibfield  {title} {\bibinfo {title} {{Planck 2018
  results. IX. Constraints on primordial non-Gaussianity}},\ }\href
  {https://doi.org/10.1051/0004-6361/201935891} {\bibfield  {journal} {\bibinfo
   {journal} {\aap}\ }\textbf {\bibinfo {volume} {641}},\ \bibinfo {eid} {A9}
  (\bibinfo {year} {2020}{\natexlab{c}})},\ \Eprint
  {https://arxiv.org/abs/1905.05697} {arXiv:1905.05697 [astro-ph.CO]}
  \BibitemShut {NoStop}%
\bibitem [{\citenamefont {{Komatsu}}\ and\ \citenamefont
  {{Spergel}}(2001)}]{2001PhRvD..63f3002K}%
  \BibitemOpen
  \bibfield  {author} {\bibinfo {author} {\bibfnamefont {E.}~\bibnamefont
  {{Komatsu}}}\ and\ \bibinfo {author} {\bibfnamefont {D.~N.}\ \bibnamefont
  {{Spergel}}},\ }\bibfield  {title} {\bibinfo {title} {{Acoustic signatures in
  the primary microwave background bispectrum}},\ }\href
  {https://doi.org/10.1103/PhysRevD.63.063002} {\bibfield  {journal} {\bibinfo
  {journal} {\prd}\ }\textbf {\bibinfo {volume} {63}},\ \bibinfo {eid} {063002}
  (\bibinfo {year} {2001})},\ \Eprint {https://arxiv.org/abs/astro-ph/0005036}
  {arXiv:astro-ph/0005036 [astro-ph]} \BibitemShut {NoStop}%
\bibitem [{\citenamefont {{Greco}}\ \emph {et~al.}(2022)\citenamefont
  {{Greco}}, \citenamefont {{Bartolo}},\ and\ \citenamefont
  {{Gruppuso}}}]{2022JCAP...03..050G}%
  \BibitemOpen
  \bibfield  {author} {\bibinfo {author} {\bibfnamefont {A.}~\bibnamefont
  {{Greco}}}, \bibinfo {author} {\bibfnamefont {N.}~\bibnamefont {{Bartolo}}},\
  and\ \bibinfo {author} {\bibfnamefont {A.}~\bibnamefont {{Gruppuso}}},\
  }\bibfield  {title} {\bibinfo {title} {{Cosmic birefrigence: cross-spectra
  and cross-bispectra with CMB anisotropies}},\ }\href
  {https://doi.org/10.1088/1475-7516/2022/03/050} {\bibfield  {journal}
  {\bibinfo  {journal} {\jcap}\ }\textbf {\bibinfo {volume} {2022}},\ \bibinfo
  {eid} {050} (\bibinfo {year} {2022})},\ \Eprint
  {https://arxiv.org/abs/2202.04584} {arXiv:2202.04584 [astro-ph.CO]}
  \BibitemShut {NoStop}%
\bibitem [{\citenamefont {{Duivenvoorden}}\ \emph {et~al.}(2020)\citenamefont
  {{Duivenvoorden}}, \citenamefont {{Meerburg}},\ and\ \citenamefont
  {{Freese}}}]{2020PhRvD.102b3521D}%
  \BibitemOpen
  \bibfield  {author} {\bibinfo {author} {\bibfnamefont {A.~J.}\ \bibnamefont
  {{Duivenvoorden}}}, \bibinfo {author} {\bibfnamefont {P.~D.}\ \bibnamefont
  {{Meerburg}}},\ and\ \bibinfo {author} {\bibfnamefont {K.}~\bibnamefont
  {{Freese}}},\ }\bibfield  {title} {\bibinfo {title} {{CMB B -mode
  non-Gaussianity: Optimal bispectrum estimator and Fisher forecasts}},\ }\href
  {https://doi.org/10.1103/PhysRevD.102.023521} {\bibfield  {journal} {\bibinfo
   {journal} {\prd}\ }\textbf {\bibinfo {volume} {102}},\ \bibinfo {eid}
  {023521} (\bibinfo {year} {2020})},\ \Eprint
  {https://arxiv.org/abs/1911.11349} {arXiv:1911.11349 [astro-ph.CO]}
  \BibitemShut {NoStop}%
\bibitem [{\citenamefont {{Shiraishi}}(2016)}]{2016PhRvD..94h3503S}%
  \BibitemOpen
  \bibfield  {author} {\bibinfo {author} {\bibfnamefont {M.}~\bibnamefont
  {{Shiraishi}}},\ }\bibfield  {title} {\bibinfo {title} {{Parity violation in
  the CMB trispectrum from the scalar sector}},\ }\href
  {https://doi.org/10.1103/PhysRevD.94.083503} {\bibfield  {journal} {\bibinfo
  {journal} {\prd}\ }\textbf {\bibinfo {volume} {94}},\ \bibinfo {eid} {083503}
  (\bibinfo {year} {2016})},\ \Eprint {https://arxiv.org/abs/1608.00368}
  {arXiv:1608.00368 [astro-ph.CO]} \BibitemShut {NoStop}%
\bibitem [{\citenamefont {{Meerburg}}\ \emph {et~al.}(2016)\citenamefont
  {{Meerburg}}, \citenamefont {{Meyers}}, \citenamefont {{van Engelen}},\ and\
  \citenamefont {{Ali-Ha{\"\i}moud}}}]{2016PhRvD..93l3511M}%
  \BibitemOpen
  \bibfield  {author} {\bibinfo {author} {\bibfnamefont {P.~D.}\ \bibnamefont
  {{Meerburg}}}, \bibinfo {author} {\bibfnamefont {J.}~\bibnamefont
  {{Meyers}}}, \bibinfo {author} {\bibfnamefont {A.}~\bibnamefont {{van
  Engelen}}},\ and\ \bibinfo {author} {\bibfnamefont {Y.}~\bibnamefont
  {{Ali-Ha{\"\i}moud}}},\ }\bibfield  {title} {\bibinfo {title} {{CMB B -mode
  non-Gaussianity}},\ }\href {https://doi.org/10.1103/PhysRevD.93.123511}
  {\bibfield  {journal} {\bibinfo  {journal} {\prd}\ }\textbf {\bibinfo
  {volume} {93}},\ \bibinfo {eid} {123511} (\bibinfo {year} {2016})},\ \Eprint
  {https://arxiv.org/abs/1603.02243} {arXiv:1603.02243 [astro-ph.CO]}
  \BibitemShut {NoStop}%
\bibitem [{\citenamefont {{Bucher}}\ \emph {et~al.}(2016)\citenamefont
  {{Bucher}}, \citenamefont {{Racine}},\ and\ \citenamefont {{van
  Tent}}}]{2016JCAP...05..055B}%
  \BibitemOpen
  \bibfield  {author} {\bibinfo {author} {\bibfnamefont {M.}~\bibnamefont
  {{Bucher}}}, \bibinfo {author} {\bibfnamefont {B.}~\bibnamefont {{Racine}}},\
  and\ \bibinfo {author} {\bibfnamefont {B.}~\bibnamefont {{van Tent}}},\
  }\bibfield  {title} {\bibinfo {title} {{The binned bispectrum estimator:
  template-based and non-parametric CMB non-Gaussianity searches}},\ }\href
  {https://doi.org/10.1088/1475-7516/2016/05/055} {\bibfield  {journal}
  {\bibinfo  {journal} {\jcap}\ }\textbf {\bibinfo {volume} {2016}},\ \bibinfo
  {eid} {055} (\bibinfo {year} {2016})},\ \Eprint
  {https://arxiv.org/abs/1509.08107} {arXiv:1509.08107 [astro-ph.CO]}
  \BibitemShut {NoStop}%
\bibitem [{\citenamefont {{Bartolo}}\ \emph {et~al.}(2016)\citenamefont
  {{Bartolo}}, \citenamefont {{Liguori}},\ and\ \citenamefont
  {{Shiraishi}}}]{2016JCAP...03..029B}%
  \BibitemOpen
  \bibfield  {author} {\bibinfo {author} {\bibfnamefont {N.}~\bibnamefont
  {{Bartolo}}}, \bibinfo {author} {\bibfnamefont {M.}~\bibnamefont
  {{Liguori}}},\ and\ \bibinfo {author} {\bibfnamefont {M.}~\bibnamefont
  {{Shiraishi}}},\ }\bibfield  {title} {\bibinfo {title} {{Primordial
  trispectra and CMB spectral distortions}},\ }\href
  {https://doi.org/10.1088/1475-7516/2016/03/029} {\bibfield  {journal}
  {\bibinfo  {journal} {\jcap}\ }\textbf {\bibinfo {volume} {2016}},\ \bibinfo
  {eid} {029} (\bibinfo {year} {2016})},\ \Eprint
  {https://arxiv.org/abs/1511.01474} {arXiv:1511.01474 [astro-ph.CO]}
  \BibitemShut {NoStop}%
\bibitem [{\citenamefont {{Luo}}\ and\ \citenamefont
  {{Schramm}}(1993{\natexlab{b}})}]{1993ApJ...408...33L}%
  \BibitemOpen
  \bibfield  {author} {\bibinfo {author} {\bibfnamefont {X.}~\bibnamefont
  {{Luo}}}\ and\ \bibinfo {author} {\bibfnamefont {D.~N.}\ \bibnamefont
  {{Schramm}}},\ }\bibfield  {title} {\bibinfo {title} {{Kurtosis, Skewness,
  and Non-Gaussian Cosmological Density Perturbations}},\ }\href
  {https://doi.org/10.1086/172567} {\bibfield  {journal} {\bibinfo  {journal}
  {\apj}\ }\textbf {\bibinfo {volume} {408}},\ \bibinfo {pages} {33} (\bibinfo
  {year} {1993}{\natexlab{b}})}\BibitemShut {NoStop}%
\bibitem [{\citenamefont {{Khan}}\ and\ \citenamefont
  {{Saha}}(2022)}]{2022JCAP...06..006K}%
  \BibitemOpen
  \bibfield  {author} {\bibinfo {author} {\bibfnamefont {M.~I.}\ \bibnamefont
  {{Khan}}}\ and\ \bibinfo {author} {\bibfnamefont {R.}~\bibnamefont
  {{Saha}}},\ }\bibfield  {title} {\bibinfo {title} {{Isotropy statistics of
  CMB hot and cold spots}},\ }\href
  {https://doi.org/10.1088/1475-7516/2022/06/006} {\bibfield  {journal}
  {\bibinfo  {journal} {\jcap}\ }\textbf {\bibinfo {volume} {2022}},\ \bibinfo
  {eid} {006} (\bibinfo {year} {2022})},\ \Eprint
  {https://arxiv.org/abs/2111.05886} {arXiv:2111.05886 [astro-ph.CO]}
  \BibitemShut {NoStop}%
\bibitem [{\citenamefont {{Novikov}}\ \emph {et~al.}(1999)\citenamefont
  {{Novikov}}, \citenamefont {{Feldman}},\ and\ \citenamefont
  {{Shandarin}}}]{1999IJMPD...8..291N}%
  \BibitemOpen
  \bibfield  {author} {\bibinfo {author} {\bibfnamefont {D.}~\bibnamefont
  {{Novikov}}}, \bibinfo {author} {\bibfnamefont {H.~A.}\ \bibnamefont
  {{Feldman}}},\ and\ \bibinfo {author} {\bibfnamefont {S.~F.}\ \bibnamefont
  {{Shandarin}}},\ }\bibfield  {title} {\bibinfo {title} {{Minkowski
  Functionals and Cluster Analysis for CMB Maps}},\ }\href
  {https://doi.org/10.1142/S0218271899000225} {\bibfield  {journal} {\bibinfo
  {journal} {International Journal of Modern Physics D}\ }\textbf {\bibinfo
  {volume} {8}},\ \bibinfo {pages} {291} (\bibinfo {year} {1999})},\ \Eprint
  {https://arxiv.org/abs/astro-ph/9809238} {arXiv:astro-ph/9809238 [astro-ph]}
  \BibitemShut {NoStop}%
\bibitem [{\citenamefont {{Barreiro}}\ \emph {et~al.}(1998)\citenamefont
  {{Barreiro}}, \citenamefont {{Sanz}}, \citenamefont {{Martinez-Gonzalez}},\
  and\ \citenamefont {{Silk}}}]{1998MNRAS.296..693B}%
  \BibitemOpen
  \bibfield  {author} {\bibinfo {author} {\bibfnamefont {R.~B.}\ \bibnamefont
  {{Barreiro}}}, \bibinfo {author} {\bibfnamefont {J.~L.}\ \bibnamefont
  {{Sanz}}}, \bibinfo {author} {\bibfnamefont {E.}~\bibnamefont
  {{Martinez-Gonzalez}}},\ and\ \bibinfo {author} {\bibfnamefont
  {J.}~\bibnamefont {{Silk}}},\ }\bibfield  {title} {\bibinfo {title}
  {{Correlation of excursion sets for non-Gaussian cosmic microwave background
  temperature distributions}},\ }\href
  {https://doi.org/10.1046/j.1365-8711.1998.01399.x} {\bibfield  {journal}
  {\bibinfo  {journal} {\mnras}\ }\textbf {\bibinfo {volume} {296}},\ \bibinfo
  {pages} {693} (\bibinfo {year} {1998})},\ \Eprint
  {https://arxiv.org/abs/astro-ph/9712143} {arXiv:astro-ph/9712143 [astro-ph]}
  \BibitemShut {NoStop}%
\bibitem [{\citenamefont {{Kotok}}\ \emph {et~al.}(2001)\citenamefont
  {{Kotok}}, \citenamefont {{Novikov}}, \citenamefont {{Naselsky}},
  \citenamefont {{Novikov}},\ and\ \citenamefont
  {{Dolgov}}}]{2001IJMPD..10..501K}%
  \BibitemOpen
  \bibfield  {author} {\bibinfo {author} {\bibfnamefont {E.}~\bibnamefont
  {{Kotok}}}, \bibinfo {author} {\bibfnamefont {I.~D.}\ \bibnamefont
  {{Novikov}}}, \bibinfo {author} {\bibfnamefont {P.}~\bibnamefont
  {{Naselsky}}}, \bibinfo {author} {\bibfnamefont {D.}~\bibnamefont
  {{Novikov}}},\ and\ \bibinfo {author} {\bibfnamefont {A.}~\bibnamefont
  {{Dolgov}}},\ }\bibfield  {title} {\bibinfo {title} {{Peculiarities of
  Anisotropy and Polarization as an Indicator of Noises in the CMB Maps}},\
  }\href {https://doi.org/10.1142/S0218271801001153} {\bibfield  {journal}
  {\bibinfo  {journal} {International Journal of Modern Physics D}\ }\textbf
  {\bibinfo {volume} {10}},\ \bibinfo {pages} {501} (\bibinfo {year} {2001})},\
  \Eprint {https://arxiv.org/abs/astro-ph/0011521} {arXiv:astro-ph/0011521
  [astro-ph]} \BibitemShut {NoStop}%
\bibitem [{\citenamefont {{Vafaei Sadr}}\ and\ \citenamefont
  {{Movahed}}(2021)}]{2021MNRAS.503..815V}%
  \BibitemOpen
  \bibfield  {author} {\bibinfo {author} {\bibfnamefont {A.}~\bibnamefont
  {{Vafaei Sadr}}}\ and\ \bibinfo {author} {\bibfnamefont {S.~M.~S.}\
  \bibnamefont {{Movahed}}},\ }\bibfield  {title} {\bibinfo {title}
  {{Clustering of local extrema in Planck CMB maps}},\ }\href
  {https://doi.org/10.1093/mnras/stab368} {\bibfield  {journal} {\bibinfo
  {journal} {\mnras}\ }\textbf {\bibinfo {volume} {503}},\ \bibinfo {pages}
  {815} (\bibinfo {year} {2021})},\ \Eprint {https://arxiv.org/abs/2003.07364}
  {arXiv:2003.07364 [astro-ph.CO]} \BibitemShut {NoStop}%
\bibitem [{\citenamefont {{Ben-David}}\ \emph {et~al.}(2015)\citenamefont
  {{Ben-David}}, \citenamefont {{Liu}},\ and\ \citenamefont
  {{Jackson}}}]{2015JCAP...06..051B}%
  \BibitemOpen
  \bibfield  {author} {\bibinfo {author} {\bibfnamefont {A.}~\bibnamefont
  {{Ben-David}}}, \bibinfo {author} {\bibfnamefont {H.}~\bibnamefont {{Liu}}},\
  and\ \bibinfo {author} {\bibfnamefont {A.~D.}\ \bibnamefont {{Jackson}}},\
  }\bibfield  {title} {\bibinfo {title} {{The Kullback-Leibler divergence as an
  estimator of the statistical properties of CMB maps}},\ }\href
  {https://doi.org/10.1088/1475-7516/2015/06/051} {\bibfield  {journal}
  {\bibinfo  {journal} {\jcap}\ }\textbf {\bibinfo {volume} {2015}},\ \bibinfo
  {pages} {051} (\bibinfo {year} {2015})},\ \Eprint
  {https://arxiv.org/abs/1506.07724} {arXiv:1506.07724 [astro-ph.CO]}
  \BibitemShut {NoStop}%
\bibitem [{\citenamefont {{Naselsky}}\ and\ \citenamefont
  {{Novikov}}(1995)}]{1995ApJ...444L...1N}%
  \BibitemOpen
  \bibfield  {author} {\bibinfo {author} {\bibfnamefont {P.~D.}\ \bibnamefont
  {{Naselsky}}}\ and\ \bibinfo {author} {\bibfnamefont {D.~I.}\ \bibnamefont
  {{Novikov}}},\ }\bibfield  {title} {\bibinfo {title} {{Percolation and
  Cluster Analysis for Delta T/T Maps}},\ }\href
  {https://doi.org/10.1086/187845} {\bibfield  {journal} {\bibinfo  {journal}
  {\apjl}\ }\textbf {\bibinfo {volume} {444}},\ \bibinfo {pages} {L1} (\bibinfo
  {year} {1995})}\BibitemShut {NoStop}%
\bibitem [{\citenamefont {{Novaes}}\ \emph {et~al.}(2015)\citenamefont
  {{Novaes}}, \citenamefont {{Bernui}}, \citenamefont {{Ferreira}},\ and\
  \citenamefont {{Wuensche}}}]{2015JCAP...09..064N}%
  \BibitemOpen
  \bibfield  {author} {\bibinfo {author} {\bibfnamefont {C.~P.}\ \bibnamefont
  {{Novaes}}}, \bibinfo {author} {\bibfnamefont {A.}~\bibnamefont {{Bernui}}},
  \bibinfo {author} {\bibfnamefont {I.~S.}\ \bibnamefont {{Ferreira}}},\ and\
  \bibinfo {author} {\bibfnamefont {C.~A.}\ \bibnamefont {{Wuensche}}},\
  }\bibfield  {title} {\bibinfo {title} {{A neural-network based estimator to
  search for primordial non-Gaussianity in Planck CMB maps}},\ }\href
  {https://doi.org/10.1088/1475-7516/2015/09/064} {\bibfield  {journal}
  {\bibinfo  {journal} {\jcap}\ }\textbf {\bibinfo {volume} {2015}},\ \bibinfo
  {pages} {064} (\bibinfo {year} {2015})},\ \Eprint
  {https://arxiv.org/abs/1409.3876} {arXiv:1409.3876 [astro-ph.CO]}
  \BibitemShut {NoStop}%
\bibitem [{\citenamefont {{Novaes}}\ \emph {et~al.}(2014)\citenamefont
  {{Novaes}}, \citenamefont {{Bernui}}, \citenamefont {{Ferreira}},\ and\
  \citenamefont {{Wuensche}}}]{2014JCAP...01..018N}%
  \BibitemOpen
  \bibfield  {author} {\bibinfo {author} {\bibfnamefont {C.~P.}\ \bibnamefont
  {{Novaes}}}, \bibinfo {author} {\bibfnamefont {A.}~\bibnamefont {{Bernui}}},
  \bibinfo {author} {\bibfnamefont {I.~S.}\ \bibnamefont {{Ferreira}}},\ and\
  \bibinfo {author} {\bibfnamefont {C.~A.}\ \bibnamefont {{Wuensche}}},\
  }\bibfield  {title} {\bibinfo {title} {{Searching for primordial
  non-Gaussianity in Planck CMB maps using a combined estimator}},\ }\href
  {https://doi.org/10.1088/1475-7516/2014/01/018} {\bibfield  {journal}
  {\bibinfo  {journal} {\jcap}\ }\textbf {\bibinfo {volume} {2014}},\ \bibinfo
  {eid} {018} (\bibinfo {year} {2014})},\ \Eprint
  {https://arxiv.org/abs/1312.3293} {arXiv:1312.3293 [astro-ph.CO]}
  \BibitemShut {NoStop}%
\bibitem [{\citenamefont {{Remazeilles}}\ and\ \citenamefont
  {{Chluba}}(2018)}]{2018MNRAS.478..807R}%
  \BibitemOpen
  \bibfield  {author} {\bibinfo {author} {\bibfnamefont {M.}~\bibnamefont
  {{Remazeilles}}}\ and\ \bibinfo {author} {\bibfnamefont {J.}~\bibnamefont
  {{Chluba}}},\ }\bibfield  {title} {\bibinfo {title} {{Extracting
  foreground-obscured {\ensuremath{\mu}}-distortion anisotropies to constrain
  primordial non-Gaussianity}},\ }\href {https://doi.org/10.1093/mnras/sty1034}
  {\bibfield  {journal} {\bibinfo  {journal} {\mnras}\ }\textbf {\bibinfo
  {volume} {478}},\ \bibinfo {pages} {807} (\bibinfo {year} {2018})},\ \Eprint
  {https://arxiv.org/abs/1802.10101} {arXiv:1802.10101 [astro-ph.CO]}
  \BibitemShut {NoStop}%
\bibitem [{\citenamefont {{Khatri}}\ and\ \citenamefont
  {{Sunyaev}}(2015)}]{2015JCAP...09..026K}%
  \BibitemOpen
  \bibfield  {author} {\bibinfo {author} {\bibfnamefont {R.}~\bibnamefont
  {{Khatri}}}\ and\ \bibinfo {author} {\bibfnamefont {R.}~\bibnamefont
  {{Sunyaev}}},\ }\bibfield  {title} {\bibinfo {title} {{Constraints on
  {\ensuremath{\mu}}-distortion fluctuations and primordial non-Gaussianity
  from Planck data}},\ }\href {https://doi.org/10.1088/1475-7516/2015/09/026}
  {\bibfield  {journal} {\bibinfo  {journal} {\jcap}\ }\textbf {\bibinfo
  {volume} {2015}},\ \bibinfo {pages} {026} (\bibinfo {year} {2015})},\ \Eprint
  {https://arxiv.org/abs/1507.05615} {arXiv:1507.05615 [astro-ph.CO]}
  \BibitemShut {NoStop}%
\bibitem [{\citenamefont {{Rotti}}\ \emph {et~al.}(2022)\citenamefont
  {{Rotti}}, \citenamefont {{Ravenni}},\ and\ \citenamefont
  {{Chluba}}}]{2022MNRAS.515.5847R}%
  \BibitemOpen
  \bibfield  {author} {\bibinfo {author} {\bibfnamefont {A.}~\bibnamefont
  {{Rotti}}}, \bibinfo {author} {\bibfnamefont {A.}~\bibnamefont {{Ravenni}}},\
  and\ \bibinfo {author} {\bibfnamefont {J.}~\bibnamefont {{Chluba}}},\
  }\bibfield  {title} {\bibinfo {title} {{Non-Gaussianity constraints with
  anisotropic {\ensuremath{\mu}} distortion measurements from Planck}},\ }\href
  {https://doi.org/10.1093/mnras/stac2082} {\bibfield  {journal} {\bibinfo
  {journal} {\mnras}\ }\textbf {\bibinfo {volume} {515}},\ \bibinfo {pages}
  {5847} (\bibinfo {year} {2022})},\ \Eprint {https://arxiv.org/abs/2205.15971}
  {arXiv:2205.15971 [astro-ph.CO]} \BibitemShut {NoStop}%
\bibitem [{\citenamefont {{Zegeye}}\ \emph {et~al.}(2023)\citenamefont
  {{Zegeye}}, \citenamefont {{Bianchini}}, \citenamefont {{Bond}},
  \citenamefont {{Chluba}}, \citenamefont {{Crawford}}, \citenamefont
  {{Fabbian}}, \citenamefont {{Gluscevic}}, \citenamefont {{Grin}},
  \citenamefont {{Hill}}, \citenamefont {{Meerburg}}, \citenamefont
  {{Orlando}}, \citenamefont {{Partridge}}, \citenamefont {{Reichardt}},
  \citenamefont {{Remazeilles}}, \citenamefont {{Scott}}, \citenamefont
  {{Wollack}},\ and\ \citenamefont {{CMB-S4
  Collaboration}}}]{2023PhRvD.108j3536Z}%
  \BibitemOpen
  \bibfield  {author} {\bibinfo {author} {\bibfnamefont {D.}~\bibnamefont
  {{Zegeye}}}, \bibinfo {author} {\bibfnamefont {F.}~\bibnamefont
  {{Bianchini}}}, \bibinfo {author} {\bibfnamefont {J.~R.}\ \bibnamefont
  {{Bond}}}, \bibinfo {author} {\bibfnamefont {J.}~\bibnamefont {{Chluba}}},
  \bibinfo {author} {\bibfnamefont {T.}~\bibnamefont {{Crawford}}}, \bibinfo
  {author} {\bibfnamefont {G.}~\bibnamefont {{Fabbian}}}, \bibinfo {author}
  {\bibfnamefont {V.}~\bibnamefont {{Gluscevic}}}, \bibinfo {author}
  {\bibfnamefont {D.}~\bibnamefont {{Grin}}}, \bibinfo {author} {\bibfnamefont
  {J.~C.}\ \bibnamefont {{Hill}}}, \bibinfo {author} {\bibfnamefont {P.~D.}\
  \bibnamefont {{Meerburg}}}, \bibinfo {author} {\bibfnamefont
  {G.}~\bibnamefont {{Orlando}}}, \bibinfo {author} {\bibfnamefont
  {B.}~\bibnamefont {{Partridge}}}, \bibinfo {author} {\bibfnamefont {C.~L.}\
  \bibnamefont {{Reichardt}}}, \bibinfo {author} {\bibfnamefont
  {M.}~\bibnamefont {{Remazeilles}}}, \bibinfo {author} {\bibfnamefont
  {D.}~\bibnamefont {{Scott}}}, \bibinfo {author} {\bibfnamefont {E.~J.}\
  \bibnamefont {{Wollack}}},\ and\ \bibinfo {author} {\bibnamefont {{CMB-S4
  Collaboration}}},\ }\bibfield  {title} {\bibinfo {title} {{CMB-S4 forecasts
  for constraints on f$_{NL}$ through {\ensuremath{\mu}} -distortion
  anisotropy}},\ }\href {https://doi.org/10.1103/PhysRevD.108.103536}
  {\bibfield  {journal} {\bibinfo  {journal} {\prd}\ }\textbf {\bibinfo
  {volume} {108}},\ \bibinfo {eid} {103536} (\bibinfo {year} {2023})},\ \Eprint
  {https://arxiv.org/abs/2303.00916} {arXiv:2303.00916 [astro-ph.CO]}
  \BibitemShut {NoStop}%
\bibitem [{\citenamefont {{Giannantonio}}\ and\ \citenamefont
  {{Percival}}(2014)}]{2014MNRAS.441L..16G}%
  \BibitemOpen
  \bibfield  {author} {\bibinfo {author} {\bibfnamefont {T.}~\bibnamefont
  {{Giannantonio}}}\ and\ \bibinfo {author} {\bibfnamefont {W.~J.}\
  \bibnamefont {{Percival}}},\ }\bibfield  {title} {\bibinfo {title} {{Using
  correlations between cosmic microwave background lensing and large-scale
  structure to measure primordial non-Gaussianity.}},\ }\href
  {https://doi.org/10.1093/mnrasl/slu036} {\bibfield  {journal} {\bibinfo
  {journal} {\mnras}\ }\textbf {\bibinfo {volume} {441}},\ \bibinfo {pages}
  {L16} (\bibinfo {year} {2014})},\ \Eprint {https://arxiv.org/abs/1312.5154}
  {arXiv:1312.5154 [astro-ph.CO]} \BibitemShut {NoStop}%
\bibitem [{\citenamefont {{Ballardini}}\ \emph {et~al.}(2019)\citenamefont
  {{Ballardini}}, \citenamefont {{Matthewson}},\ and\ \citenamefont
  {{Maartens}}}]{2019MNRAS.489.1950B}%
  \BibitemOpen
  \bibfield  {author} {\bibinfo {author} {\bibfnamefont {M.}~\bibnamefont
  {{Ballardini}}}, \bibinfo {author} {\bibfnamefont {W.~L.}\ \bibnamefont
  {{Matthewson}}},\ and\ \bibinfo {author} {\bibfnamefont {R.}~\bibnamefont
  {{Maartens}}},\ }\bibfield  {title} {\bibinfo {title} {{Constraining
  primordial non-Gaussianity using two galaxy surveys and CMB lensing}},\
  }\href {https://doi.org/10.1093/mnras/stz2258} {\bibfield  {journal}
  {\bibinfo  {journal} {\mnras}\ }\textbf {\bibinfo {volume} {489}},\ \bibinfo
  {pages} {1950} (\bibinfo {year} {2019})},\ \Eprint
  {https://arxiv.org/abs/1906.04730} {arXiv:1906.04730 [astro-ph.CO]}
  \BibitemShut {NoStop}%
\bibitem [{\citenamefont {{Dolgov}}\ \emph {et~al.}(1999)\citenamefont
  {{Dolgov}}, \citenamefont {{Doroshkevich}}, \citenamefont {{Novikov}},\ and\
  \citenamefont {{Novikov}}}]{1999IJMPD...8..189D}%
  \BibitemOpen
  \bibfield  {author} {\bibinfo {author} {\bibfnamefont {A.~D.}\ \bibnamefont
  {{Dolgov}}}, \bibinfo {author} {\bibfnamefont {A.~G.}\ \bibnamefont
  {{Doroshkevich}}}, \bibinfo {author} {\bibfnamefont {D.~I.}\ \bibnamefont
  {{Novikov}}},\ and\ \bibinfo {author} {\bibfnamefont {I.~D.}\ \bibnamefont
  {{Novikov}}},\ }\bibfield  {title} {\bibinfo {title} {{Geometry and
  Statistics of Cosmic Microwave Polarization}},\ }\href
  {https://doi.org/10.1142/S0218271899000171} {\bibfield  {journal} {\bibinfo
  {journal} {International Journal of Modern Physics D}\ }\textbf {\bibinfo
  {volume} {8}},\ \bibinfo {pages} {189} (\bibinfo {year} {1999})},\ \Eprint
  {https://arxiv.org/abs/astro-ph/9901399} {arXiv:astro-ph/9901399 [astro-ph]}
  \BibitemShut {NoStop}%
\bibitem [{\citenamefont {{Dolgov}}\ \emph {et~al.}(2000)\citenamefont
  {{Dolgov}}, \citenamefont {{Doroshkevich}}, \citenamefont {{Novikov}},\ and\
  \citenamefont {{Novikov}}}]{2000A&AT...19..213D}%
  \BibitemOpen
  \bibfield  {author} {\bibinfo {author} {\bibfnamefont {A.~D.}\ \bibnamefont
  {{Dolgov}}}, \bibinfo {author} {\bibfnamefont {A.~G.}\ \bibnamefont
  {{Doroshkevich}}}, \bibinfo {author} {\bibfnamefont {D.~I.}\ \bibnamefont
  {{Novikov}}},\ and\ \bibinfo {author} {\bibfnamefont {I.~D.}\ \bibnamefont
  {{Novikov}}},\ }\bibfield  {title} {\bibinfo {title} {{Geometrical methods of
  analysis of polarization of CMB}},\ }\href
  {https://doi.org/10.1080/10556790008238573} {\bibfield  {journal} {\bibinfo
  {journal} {Astronomical and Astrophysical Transactions}\ }\textbf {\bibinfo
  {volume} {19}},\ \bibinfo {pages} {213} (\bibinfo {year} {2000})}\BibitemShut
  {NoStop}%
\bibitem [{\citenamefont {{Kasak}}\ \emph {et~al.}(2021)\citenamefont
  {{Kasak}}, \citenamefont {{Creswell}}, \citenamefont {{Naselsky}},\ and\
  \citenamefont {{Liu}}}]{2021PhRvD.104b3502K}%
  \BibitemOpen
  \bibfield  {author} {\bibinfo {author} {\bibfnamefont {J.}~\bibnamefont
  {{Kasak}}}, \bibinfo {author} {\bibfnamefont {J.}~\bibnamefont {{Creswell}}},
  \bibinfo {author} {\bibfnamefont {P.}~\bibnamefont {{Naselsky}}},\ and\
  \bibinfo {author} {\bibfnamefont {H.}~\bibnamefont {{Liu}}},\ }\bibfield
  {title} {\bibinfo {title} {{Statistics of nonpolarized points in the CMB
  polarization maps}},\ }\href {https://doi.org/10.1103/PhysRevD.104.023502}
  {\bibfield  {journal} {\bibinfo  {journal} {\prd}\ }\textbf {\bibinfo
  {volume} {104}},\ \bibinfo {eid} {023502} (\bibinfo {year} {2021})},\ \Eprint
  {https://arxiv.org/abs/2012.15811} {arXiv:2012.15811 [astro-ph.CO]}
  \BibitemShut {NoStop}%
\bibitem [{\citenamefont {Zaldarriaga}\ and\ \citenamefont
  {Seljak}(1997)}]{PhysRevD.55.1830}%
  \BibitemOpen
  \bibfield  {author} {\bibinfo {author} {\bibfnamefont {M.}~\bibnamefont
  {Zaldarriaga}}\ and\ \bibinfo {author} {\bibfnamefont {U.~c.~v.}\
  \bibnamefont {Seljak}},\ }\bibfield  {title} {\bibinfo {title} {All-sky
  analysis of polarization in the microwave background},\ }\href
  {https://doi.org/10.1103/PhysRevD.55.1830} {\bibfield  {journal} {\bibinfo
  {journal} {Phys. Rev. D}\ }\textbf {\bibinfo {volume} {55}},\ \bibinfo
  {pages} {1830} (\bibinfo {year} {1997})}\BibitemShut {NoStop}%
\bibitem [{\citenamefont {{Planck Collaboration}}\ \emph
  {et~al.}(2020{\natexlab{d}})\citenamefont {{Planck Collaboration}},
  \citenamefont {{Akrami}}, \citenamefont {{Ashdown}}, \citenamefont
  {{Aumont}}, \citenamefont {{Baccigalupi}}, \citenamefont {{Ballardini}},
  \citenamefont {{Banday}}, \citenamefont {{Barreiro}}, \citenamefont
  {{Bartolo}},\ and\ \citenamefont {{Basak}}}]{2020A&A...641A...4P}%
  \BibitemOpen
  \bibfield  {author} {\bibinfo {author} {\bibnamefont {{Planck
  Collaboration}}}, \bibinfo {author} {\bibfnamefont {Y.}~\bibnamefont
  {{Akrami}}}, \bibinfo {author} {\bibfnamefont {M.}~\bibnamefont {{Ashdown}}},
  \bibinfo {author} {\bibfnamefont {J.}~\bibnamefont {{Aumont}}}, \bibinfo
  {author} {\bibfnamefont {C.}~\bibnamefont {{Baccigalupi}}}, \bibinfo {author}
  {\bibfnamefont {M.}~\bibnamefont {{Ballardini}}}, \bibinfo {author}
  {\bibfnamefont {A.~J.}\ \bibnamefont {{Banday}}}, \bibinfo {author}
  {\bibfnamefont {R.~B.}\ \bibnamefont {{Barreiro}}}, \bibinfo {author}
  {\bibfnamefont {N.}~\bibnamefont {{Bartolo}}},\ and\ \bibinfo {author}
  {\bibfnamefont {S.~e.~a.}\ \bibnamefont {{Basak}}},\ }\bibfield  {title}
  {\bibinfo {title} {{Planck 2018 results. IV. Diffuse component separation}},\
  }\href {https://doi.org/10.1051/0004-6361/201833881} {\bibfield  {journal}
  {\bibinfo  {journal} {\aap}\ }\textbf {\bibinfo {volume} {641}},\ \bibinfo
  {eid} {A4} (\bibinfo {year} {2020}{\natexlab{d}})},\ \Eprint
  {https://arxiv.org/abs/1807.06208} {arXiv:1807.06208 [astro-ph.CO]}
  \BibitemShut {NoStop}%
\bibitem [{\citenamefont {{G{\'o}rski}}\ \emph {et~al.}(2005)\citenamefont
  {{G{\'o}rski}}, \citenamefont {{Hivon}}, \citenamefont {{Banday}},
  \citenamefont {{Wandelt}}, \citenamefont {{Hansen}}, \citenamefont
  {{Reinecke}},\ and\ \citenamefont {{Bartelmann}}}]{2005ApJ...622..759G}%
  \BibitemOpen
  \bibfield  {author} {\bibinfo {author} {\bibfnamefont {K.~M.}\ \bibnamefont
  {{G{\'o}rski}}}, \bibinfo {author} {\bibfnamefont {E.}~\bibnamefont
  {{Hivon}}}, \bibinfo {author} {\bibfnamefont {A.~J.}\ \bibnamefont
  {{Banday}}}, \bibinfo {author} {\bibfnamefont {B.~D.}\ \bibnamefont
  {{Wandelt}}}, \bibinfo {author} {\bibfnamefont {F.~K.}\ \bibnamefont
  {{Hansen}}}, \bibinfo {author} {\bibfnamefont {M.}~\bibnamefont
  {{Reinecke}}},\ and\ \bibinfo {author} {\bibfnamefont {M.}~\bibnamefont
  {{Bartelmann}}},\ }\bibfield  {title} {\bibinfo {title} {{HEALPix: A
  Framework for High-Resolution Discretization and Fast Analysis of Data
  Distributed on the Sphere}},\ }\href {https://doi.org/10.1086/427976}
  {\bibfield  {journal} {\bibinfo  {journal} {\apj}\ }\textbf {\bibinfo
  {volume} {622}},\ \bibinfo {pages} {759} (\bibinfo {year} {2005})},\ \Eprint
  {https://arxiv.org/abs/astro-ph/0409513} {arXiv:astro-ph/0409513 [astro-ph]}
  \BibitemShut {NoStop}%
\bibitem [{\citenamefont {{Planck Collaboration}}\ \emph
  {et~al.}(2016)\citenamefont {{Planck Collaboration}}, \citenamefont {{Adam}},
  \citenamefont {{Ade}}, \citenamefont {{Aghanim}}, \citenamefont {{Alves}},
  \citenamefont {{Arnaud}}, \citenamefont {{Ashdown}}, \citenamefont
  {{Aumont}}, \citenamefont {{Baccigalupi}},\ and\ \citenamefont
  {{Banday}}}]{2016A&A...594A..10P}%
  \BibitemOpen
  \bibfield  {author} {\bibinfo {author} {\bibnamefont {{Planck
  Collaboration}}}, \bibinfo {author} {\bibfnamefont {R.}~\bibnamefont
  {{Adam}}}, \bibinfo {author} {\bibfnamefont {P.~A.~R.}\ \bibnamefont
  {{Ade}}}, \bibinfo {author} {\bibfnamefont {N.}~\bibnamefont {{Aghanim}}},
  \bibinfo {author} {\bibfnamefont {M.~I.~R.}\ \bibnamefont {{Alves}}},
  \bibinfo {author} {\bibfnamefont {M.}~\bibnamefont {{Arnaud}}}, \bibinfo
  {author} {\bibfnamefont {M.}~\bibnamefont {{Ashdown}}}, \bibinfo {author}
  {\bibfnamefont {J.}~\bibnamefont {{Aumont}}}, \bibinfo {author}
  {\bibfnamefont {C.}~\bibnamefont {{Baccigalupi}}},\ and\ \bibinfo {author}
  {\bibfnamefont {A.~J. e.~a.}\ \bibnamefont {{Banday}}},\ }\bibfield  {title}
  {\bibinfo {title} {{Planck 2015 results. X. Diffuse component separation:
  Foreground maps}},\ }\href {https://doi.org/10.1051/0004-6361/201525967}
  {\bibfield  {journal} {\bibinfo  {journal} {\aap}\ }\textbf {\bibinfo
  {volume} {594}},\ \bibinfo {eid} {A10} (\bibinfo {year} {2016})},\ \Eprint
  {https://arxiv.org/abs/1502.01588} {arXiv:1502.01588 [astro-ph.CO]}
  \BibitemShut {NoStop}%
\end{thebibliography}%




\end{document}